\documentclass[useAMS,usenatbib]{mn2e}
\usepackage{aasabbrev}
\usepackage{graphicx}

\defcitealias{ciotti:07}{C07} 	  	 

\newcommand{\unit}[1]{\,\hbox{#1}}

\newcommand{\ithree}{\ensuremath{Q}}
\newcommand{\mi}{\ensuremath{J}}

\newcommand{\eqnref}[1]{eq. (\ref{#1})}

\newcommand{\eqntworef}[2]{eqs. (\ref{#1}) and (\ref{#2})}
\newcommand{\timescale}{t}
\newcommand{\opticaldepth}{\tau}
\newcommand{\pabs}{p_{\rm abs}}
\newcommand{\tildepabs}{\tilde{p}_{\rm abs}}
\newcommand{\ptrans}{p_{\rm trans }}
\newcommand{\flux}{F}
\newcommand{\force}{f}
\newcommand{\thetapos}{\tilde{\theta}}
\newcommand{\phipos}{\tilde{\phi}}
\newcommand{\thetaprop}{\theta}
\newcommand{\phiprop}{\phi}
\title[Radiative Transfer and Radiative driving of Outflows in AGN and
Starbursts]{Radiative Transfer and Radiative driving of Outflows in AGN and
Starbursts}
\author[G. S. Novak, J. P. Ostriker, and L. Ciotti]{G. S. Novak$^{1,2}$\thanks{E-mail:
novak@astro.princeton.edu (GSN)}, J. P. Ostriker$^{1}$, and L. Ciotti$^{3}$\\
$^{1}$Department of Astrophysical Sciences, Peyton Hall, Princeton University, Princeton, NJ 08544, United States\\
$^{1}$Paris Observatory, 61 Avenue de l'Observatoire, 75014 Paris, France \\
$^{3}$Department of Astronomy, University of Bologna, via Ranzani 1, I-40127, Bologna, Italy}
\begin{document}

\date{Accepted 2012 Jan 1. Received 2012 Jan 1; in original form 2012 Jan 1}

\pagerange{\pageref{firstpage}--\pageref{lastpage}} \pubyear{2012}

\maketitle

\label{firstpage}

\begin{abstract}
  To facilitate the study of black hole fueling, star formation, and
  feedback in galaxies, we outline a method for treating the radial
  forces on interstellar gas due to absorption of photons by dust
  grains.  The method gives the correct behavior in all of the
  relevant limits (dominated by the central point source; dominated by
  the distributed isotropic source; optically thin; optically thick to
  UV/optical; optically thick to IR) and reasonably interpolates
  between the limits when necessary.  The method is explicitly energy
  conserving so that UV/optical photons that are absorbed are not
  lost, but are rather redistributed to the IR where they may scatter
  out of the galaxy.  We implement the radiative transfer algorithm in
  a two-dimensional hydrodynamical code designed to study feedback
  processes in the context of early-type galaxies.  We find that the
  dynamics and final state of simulations are measurably but only
  moderately affected by radiative forces on dust, even when
  assumptions about the dust-to-gas ratio are varied from zero to a
  value appropriate for the Milky Way.  In simulations with high gas
  densities designed to mimic ULIRGs with a star formation rate of
  several hundred solar masses per year, dust makes a more substantial
  contribution to the dynamics and outcome of the simulation.  We find
  that, despite the large opacity of dust to UV radiation, the
  momentum input to the flow from radiation very rarely exceeds L/c
  due to two factors: the low opacity of dust to the re-radiated IR
  and the tendency for dust to be destroyed by sputtering in hot gas
  environments.  We also develop a simplification of our radiative
  transfer algorithm that respects the essential physics but is much
  easier to implement and requires a fraction of the computational
  cost.
\end{abstract}


\section{Introduction}

Supermassive black holes (SMBHs) are thought to inhabit the central
regions of nearly all
massive galaxies and have properties well-correlated with those of their
host galaxies \citep{gebhardt:00, ferrarese:00, tremaine:02,
  novak:06-blackholes, gueltekin:09}.  The direction of the causal
link between SMBH and host galaxy properties remains unclear, but the
existence of such a correlation implies an intimate connection between
the physics of SMBH growth and galaxy formation, spanning a factor of
at least $10^4$ in length scale from $\sim 1 \unit{pc}$ to $\sim 10
\unit{kpc}$.

Quasars came to astronomers' attention due to their enormous
electromagnetic output.  Bright active galactic nuclei (AGN) emit as
much as $10^{47} \unit{erg} \unit{s}^{-1}$ over the electromagnetic spectrum
\citep{woo:02, kollmeier:06}.  A significant amount of careful work
has been performed to study the physical effects of this extraordinary
outpouring of electromagnetic energy and momentum on the material in
the immediate vicinity of the AGN \citep[e.g.][]{proga:00, proga:08,
  kurosawa:09-3d, park:11, park:12}.  As a result we are beginning to understand the
radiation driven broad-line winds emerging from discs around the
central SMBHs. However, relatively little work has been done regarding
the radiative effects on the surrounding galaxy.

There has been intense theoretical interest in the ability of AGN to
affect the star formation histories and observed colors of galaxies
\citep[e.g.][]{croton:06, schawinski:07}.  The relationships between
star formation, and AGN activity, galaxy mergers, and secular
processes have turned out to be complex and the parameter space is not
yet fully mapped.  What is clear is that gas added to the central
regions of galaxies will induce a central star burst and an AGN
flare-up, but the triggering mechanism remains uncertain.

Although the use of numerical simulations to study the interactions
between black holes and galaxies has a long history
\citep[e.g.][]{binney:95, ciotti:97}, work by \citet{dimatteo:05} and
\citet{springel:05-blackholes} set off an explosion of theoretical
work.  They studied the simplest reasonable fully specified model of black
hole fueling and AGN feedback in the context of a major mergers of
gas-rich spiral galaxies and found that the AGN has a profound effect
on the gas dynamics and star formation history in the merger remnant.

During the combined AGN/starburst phase, analytical investigations by
\citet{thompson:05} and \citet{murray:05} have shown the importance of
radiative forces on dust.  Subsequent numerical work
\citep{debuhr:10,debuhr:11,debuhr:12} has verified that if the central $\simeq$
100 pc of galaxies are assumed to have optical depths of 10 to 25 in
the infrared, due to dust opacity, then radiative forces on dust have
dramatic dynamical consequences in the context of merging galaxies.

As the investigations have become more detailed and accurate, it has
become increasingly necessary to perform the radiative computations in
a more rigorous fashion, carefully determining both the energy and
momentum input into the ambient gas.  It is the purpose of this paper
to outline some of the principles and techniques that may be used for
this task.

Our goal is to develop a numerical method to self-consistently solve
the radiative transfer problem in galaxies in the presence of dust and
radiation produced by stars and AGN.  That is, we would like to solve
for the specific intensity of the radiation field for a given
distribution of radiation sources in the presence of scattering and
absorbing media, and subsequently use the specific intensity to
compute the forces on the scattering/absorbing media.  Studies of SMBH
fueling and AGN feedback performed to date have either ignored
radiative effects or else assumed a particular form of the photon
field which is then used to compute forces on the gas.

Although it would be ideal to obtain a perfect solution for the
radiation field in all cases, we will settle for a method that gives
the correct asymptotic behavior in all relevant limits and
interpolates reasonably between the limits.  It should respect global
conservation laws for energy and momentum.  It must not be
overwhelmingly demanding in terms of algorithmic complexity or
computational resources.

The paper is organized as follows: In
Section \ref{sec:previous-work} we summarize previous theoretical
work on AGN feedback as well as some recent observational results.  In
Section \ref{sec:radiative-transfer} we discuss the radiative
transport equation in the context of the present problem.  We describe
the physics behind it, its most important properties, and the
different approaches that can be used for the numerical solution.  In
Section \ref{sec:hydro-simulations} we describe the physics
implemented in a two-dimensional hydrodynamical code designed to study
black hole fueling and radiative feedback from AGN and star formation
in the context of early-type galaxies.  In Section \ref{sec:results}
we present the main results of the simulations.  Finally in Section
\ref{sec:discussion} we summarize our conclusions.

\section{Previous Work}
\label{sec:previous-work}

It has long been known that the bulk of the matter that enters black
holes over the history of the universe does so via radiatively
efficient accretion \citep{soltan:82,yu:02}. The resulting photons are the
clearest observational indication that SMBH accretion is taking place.

There is a prodigious amount of energy and momentum available during
Eddington-limited bursts consequent to accretion, certainly enough to
have a profound effect on the structure and dynamics of the
interstellar medium.

\subsection{Mechanical feedback}
Many AGN are observed to have broad absorption lines indicating strong
outflows.  Several physical processes operating near the SMBH could
plausibly launch these winds \citep[e.g.][]{proga:00} and they go
under the heading of mechanical feedback regardless of the details of
their origin \citep{ostriker:10}.  They are thought to be somewhat
collimated and are thus only observed in 1/4 to 1/2 of all bright AGN.
Recently, quantitative estimates of the kinetic energy in these
outflows have become available \citep{moe:09, arav:12-preprint} and
indicate that they contain 0.1-1\% of the energy in the radiative
component of the AGN output.  

This mechanical energy can couple to the ISM much more efficiently than the AGN
photons, making this an attractive mechanism for AGN feedback.
However, there are significant observational uncertainties in these
kinetic wind energy estimates, the studies have only been carried out
for a few objects, and there is no global constraint analogous to the
\citet{soltan:82} argument that allows an estimate of the global mean
efficiency of energy conversion to kinetic winds in AGN.  If the
\citet{moe:09} estimate of the kinetic luminosity is correct and we
assume all of the mechanical energy and momentum is effectively
deposited into the ISM, then mechanical feedback will dominate over
radiative feedback for column densities less than $10^{-21}$ to
$10^{-22} \unit{g}\unit{cm}^{-2}$.

\subsubsection{Simulations of mechanical feedback}
Many groups have carried out substantially similar studies of AGN
feedback in the context of galaxy mergers where the black hole
accretion rate is given by the smaller of the estimated Bondi rate and the
Eddington rate, while the AGN feedback is assumed to be purely thermal
energy injected in the vicinity of the black hole.  This is the basic
model developed by \citet{dimatteo:05} and
\citet{springel:05-blackholes} and applied extensively to compare with
observations of the quasar luminosity function, galaxy colors, and
more 
\citep[e.g.][]{hopkins:05, hopkins:06, hopkins:08-redgals}.  
\nocite{hopkins:08-qsos}
The same basic model was also used by \citet{johansson:09}, and all
authors found that AGN feedback has a dramatic effect on galaxies
provided that 5\% of the bolometric AGN output is transformed into
thermal energy within 100 pc of the black hole by mechanical winds
operating on smaller scales.  

However, observations of gas dynamics near AGN have indicated that the
actual efficiencies are a factor of 5 to 50 smaller than the value
usually used in many of these simulations \citep{moe:09,
  arav:12-preprint}.  \citet{booth:09} and \citet{choi:12-preprint}
have carried out careful studies of the underlying numerical and
physical assumptions that go into this feedback model, the latter
paper finding that including the radial momentum of the AGN ejecta in
the calculation dramatically increases the efficiency of mechanical
feedback.

\citet{dubois:10} developed a model of AGN involving anisotropic
injection of kinetic energy on scales similar to the resolution of the
simulation.  They implemented this model in the
adaptive-mesh-refinement code RAMSES \citep{teyssier:02} and studied
the implications of this feedback model on the properties of black
holes and galaxy clusters in cosmological simulations a resolution of
approximately one kiloparsec.

\citet{gaspari:12} performed simulations of the evolution of the ISM
of an elliptical galaxy in the presence of AGN feedback similar to
that of \citet{dubois:10}.  In terms of the galaxy model and the
treatment of stellar evolution and mass loss, the simulations are
substantially similar to those presented here.  The differences
include the fact that the \citet{gaspari:12} simulations are three
dimensional rather than two dimensional, have much lower spatial
resolution ($\simeq 150 \unit{pc}$ rather than $0.25 \unit{pc}$, and
include feedback {\em only} via anisotropic injection of kinetic
energy, rather than the more comprehensive feedback model presented
here.  \citet{gaspari:12} found that mechanical feedback efficiencies
in the range of those observed by \citet{moe:09} and \citet{arav:12-preprint}
were sufficient to limit black hole growth, balance atomic cooling,
and lead to observationally reasonable gas density and temperature
profiles, in substantial agreement with the conclusions of
\citet{novak:11}, \citet{ciotti:10}, and other studies.

The observational studies of \citet{moe:09} and
\citet{arav:12-preprint} indicate that this form of mechanical
feedback only uses a small fraction of the available energy.  This
motivates the study of other forms of radiative feedback.

\subsection{Radiative momentum feedback}
It is well-known that the photons couple to the ISM weakly if only the small
Thompson cross section is considered \citep[e.g.][]{ciotti:01}.  
Of the momentum available, only
a fraction $\opticaldepth_{\rm ES}$ (the optical depth to electron
scattering) couples to the ISM, while there is no additional energy
transfer to the gas in the case of pure Thompson scattering of photons
having $E \ll 1 \unit{MeV}$.  Only a small fraction of AGN are known
to be optically thick to Compton scattering, although studies of the
X-ray background indicate that Compton-thick AGN may be common at high
redshift \citep{daddi:07}.  X-rays couple more strongly than lower
energy photons \citep{sazonov:05} due to both inverse Compton effects
and coupling with resonant metal lines.  Nevertheless, the $M_{\rm BH}--\sigma$
relation \citep{gebhardt:00, ferrarese:00} has been construed as
evidence that black holes regulate their growth via momentum-driven
winds \citep{king:03}.

Radiation also couples to the ISM through scattering/absorption by
dust grains as well as atomic resonance lines.  The cross section for
dust absorption in the UV for solar metallicity gas with a
dust/gas ratio typical of the Milky Way can be 3500 times the electron
scattering cross section \citep{draine:03}, while that for resonance
lines can be the same or greater \citep{proga:00, sazonov:05}.  This
means that radiative feedback due to line or dust absorption can
provide more energy and momentum than mechanical feedback for column
densities as low as $10^{-18}$ to $10^{-19} \unit{g}\unit{cm}^{-2}$.

The potential importance of radiative momentum feedback has been known
for some time \citep[][hereafter \citetalias{ciotti:07}]{haehnelt:95,ciotti:07}, 
but the particular importance of
dust in the context of AGN and starbursts was pointed out by
\citet{murray:05} and \citet{thompson:05}.  Simulations by
\citet{debuhr:10,debuhr:11,debuhr:12} 
model the effect of
radiation pressure acting on dust at radii large compared to the black
hole but small compared to the galaxy.  Note that while the physical process
mediating the interaction between the SMBH and the ISM is different
from the mechanical model of \citet{dimatteo:05}, the photon momentum
is converted to hydrodynamic motion on scales smaller than the
resolution of the simulation ($\sim 100$ pc).  

\subsubsection{Simulations of radiative feedback}
\citet{sazonov:05} investigated the important effects of
X-ray heating and \citeauthor{ciotti:07} and collaborators have
incorporated all of the standard electromagnetic processes into their
one-dimensional treatments of the cooling flow initiated AGN outbursts
(\citetalias{ciotti:07}; \citealp{ciotti:09-feedback, ciotti:10, ciotti:12}).  
These investigations represent a different viewpoint from the
``conventional'' view that AGN are associated with mergers.  Here, the
AGN activity attributed to the reprocessing of gas lost by the
evolving stellar population of an isolated and ``passively'' evolving
early-type galaxy.  In section \ref{sec:observations} we comment on
recent observations that bear on this point. Recently
\citet{hensley:12-preprint} have added an improved treatment of dust
creation and destruction as well as a calculation of the dust
temperature to these one-dimensional models.

Feedback in the \citeauthor{debuhr:10} simulations is implemented via
a force on the gas assuming that there is a fixed optical depth to
IR photon scattering by dust at all times ($\opticaldepth \simeq 10$).  
No self-consistent solution
for the radiation field is sought.  
The radiative force is applied over one resolution element
near the black hole, so, although the physical model posits a radiative
origin for the computed forces, in implementation their algorithm
bears some resemblance to mechanical feedback schemes in that it is
purely local and the photons do not exert forces over macroscopic
distances in the simulation.  

\citet{novak:11} used the same physical model described in
\citet{sazonov:05} and \citetalias{ciotti:07} (with the exception of
the radiative forces on dust) to run two-dimensional simulations of SMBH
fueling and AGN feedback.  The present work extends \citet{novak:11}
to include radiative forces on dust grains.  We seek to carry out the
radiative transfer calculation due to photon absorption and scattering
by dust in a self-consistent fashion.  

\citet{hambrick:11} and \citet{kim:11} carried out simulations of
black hole fueling and AGN feedback during galaxy mergers and found
that the addition of X-ray heating by the AGN allowed the black hole
to regulate its own growth by keeping its immediate vicinity hot
without necessarily heating the majority of the gas in the galaxy.
\citet{hambrick:11} used cosmological smoothed-particle-hydrodynamics
(SPH) simulations with a simple treatment of X-ray photons from the
AGN.  \citet{kim:11} performed adaptive-mesh simulations of galaxy
mergers reaching a resolution of 15 pc.  The simulations tracked
mechanical feedback in the form of a collimated outflow generated near
the black hole as well as radiative feedback in the form of X-rays
generated by the AGN.  The X-rays transferred energy and momentum to
the surrounding gas, via by a three-dimensional Monte Carlo ray
tracing algorithm.

\subsection{Angular momentum transport}
Several efforts have focused on understanding angular momentum
transport in the absence of feedback.  This eases one of the severe
computational constraints in the problem, since it is feedback that
generates high gas temperatures and large gas velocities,
necessitating very short simulation time steps and therefore very large
computational costs per simulation.  \citet{levine:08} used
adaptive-mesh cosmological simulations to examine gas transport from
cosmological scales to $\simeq$ 1 pc from the black hole.
\citet{hopkins:10-simulation} carried out nested zoom-in simulations
of the central regions of a galaxy undergoing a major merger.  Both
groups concluded that the mutual gravitational torques exerted by
clumps of gas were sufficient to transfer angular momentum away from
the central regions and permit accretion, and that the scale-free
nature of gravity allowed the process to proceed in a nested fashion
through all scales to the presumed black hole accretion disk.  It is
very attractive to split the problem of black hole fueling and AGN
feedback into a fueling part and a feedback part.  However, one would
expect gas clumpiness to be dramatically reduced if AGN are able to
heat the ISM significantly, which would in turn shut down the angular
momentum transfer process that these authors envision.  If the gas
clumps are shielded from the AGN photons by a central dusty torus, or
if the clumps of gas are self-shielding, then the process may continue
to operate in spite of AGN feedback.

\subsection{Sub-parsec simulations}
As high-resolution simulation efforts that resolve the Bondi radius
become more common, technical details of the numerical solution become
important.  \cite{barai:11} have performed a detailed study of the
numerical and physical aspects of simulating Bondi accretion using SPH
in the presence of heating by a central black hole over scales of 0.1
to 200 pc.  They conclude that typical formulations of the artificial
viscosity term typically used in SPH can cause excessive heating near
the inner boundary of the simulation.  

\citet{dorodnitsyn:12} 
have studied the
radiation hydrodynamics of the inner few parsecs of galaxies in the
presence of dust in the the flux-limited diffusion approximation.
They show that the dusty torus around AGN is plausibly supported by
radiation pressure on dust in an outflowing wind.  They are interested
in smaller physical scales than presently concern us, and they are
only interested in the limit of large optical depths to scattering by
dust in the infrared.  However, there is significant overlap in the
basic physics between their work and the present paper.

\subsection{Observations concerning mergers, star formation, and AGN}
\label{sec:observations}
The observational situation regarding the links between black hole
growth, star formation, and mergers has been somewhat confused,
although it seems that a preponderance of evidence is now indicating
that, although mergers may trigger AGN, the majority of AGN are {\em
  not} triggered by mergers.

\citet{pierce:07} used a combination of X-ray data, space-based
optical data, and ground-based optical data, all part of the AEGIS
survey \citep{davis:07-abridged}, to conclude that X-ray selected AGN
preferentially reside in early-type galaxies and that although the
X-ray selected AGN were more often members of close pairs, the
companion was usually undisturbed.  This was interpreted as evidence
against interactions triggering AGN.  

Using Sloan Digital Sky Survey data, \citet{cisternas:11} found a
strong correlation between close galaxy pairs and galactic star
formation, but {\em no} correlation between between close pairs and
AGN activity.  \citet{schawinski:10} conducted a similar study and
found superficially similar correlations, but based on their results
connecting galaxy color to stellar population age
\citep{schawinski:07} argued that the lack of AGN with close
companions was due to a delay between the final merger and the onset
of AGN activity.

\citet{ellison:11} have found that close pairs tend to have AGN
activity, but that AGN do not have an elevated close companion rate.
They argue that mergers do indeed cause AGN activity, but that the
majority of AGN are caused by other, presumably secular processes.

\citet{diamond-stanic:12} found that AGN activity is
correlated with {\em nuclear} star formation (within $\sim$100 pc),
but {\em not} with {\em global} star formation on kiloparsec scales.

\citet{rosario:11} used space-based optical and x-ray data taken as
part of the CANDELS survey \citep{grogin:11-abridged} to conclude that
there is little difference between X-ray selected AGN and quiescent
galaxies in terms of colors and stellar populations out to a redshift
of three, arguing against a scenario where a significant fraction of
AGN are triggered by mergers.  As part of the same survey,
\citet{kocevski:12-abridged} concluded that X-ray selected
AGN are not morphologically different from a mass-matched sample of
quiescent galaxies.

\section[]{The Radiative Transfer Equation}
\label{sec:radiative-transfer}
In this section, we develop a method of treating the radiative forces
forces on the ISM gas in a galaxy.  Our goal is to arrive at an
algorithm that gives the correct behavior in the many asymptotic
limits required by the physical problem at hand.

First we discuss the astrophysical requirements that motivate some of
the choices we have made in defining the method.  In the present work
we are primarily concerned with radiative forces on dust grains, but
the method works for any source of isotropic scattering or absorption
opacity.

More precisely, we derive three related algorithms for solving the
radiative transfer equation.  All of them arise from developing a set
of moment equations for the specific intensity.  For each method, we
derive set of differential equations, the solution of which give the
mean intensity and net flux of the photon field.

The first method (Section \ref{sec:mathematical-treatment}) gives a
boundary-value problem where the physical content of the set of
differential equations is quite transparent.  However, closing the set
of differential equations in this method requires us to rely rather
heavily on our physical intuition about the solution.

The second method (Appendix \ref{sec:complex-rxfer}) also gives
boundary-value problem and is more aesthetically pleasing in a formal
mathematical sense because the closure relation is derived along
with the set of differential equations---it does not require a
physically motivated guess.  The cost an additional differential
equation to solve and the fact that the physical content of the
equations is somewhat more opaque compared to the first method.

The third method (Appendix \ref{sec:simplified-rxfer}) is an attempt to
capture the essential behavior of the system while reformulating it as
an initial value problem.  The reason for this is that it is vastly
simpler to efficiently and robustly obtain the numerical solution to an
initial value problem.

Note that in this work we use the first of these three methods and present the
other two for reference and for future work.  
All three of the algorithms are self-consistent in the sense that we
first solve
for the photon field and then learn from this solution whether the
radiation field is nearly isotropic (like the interior of a star) or
highly directed (like a point source), whether the energy is carried
by UV, Optical, or IR photons, and so on.  We do not assume a priori
that the system is in any particular asymptotic limit of the radiative
transfer equation.

We neglect angular force terms and although we use spherical symmetry
to derive the differential equations, we argue in Section
\ref{sec:ray-by-ray} that the
method can be used with good results even in the case where the system
is not spherically symmetric.

\subsection{Astrophysical preliminaries}
\label{sec:astrophysical-preliminaries}

Over their lifetimes, galaxies explore essentially all of the
asymptotic limits of the radiative transfer equation.  Most of the
time, most lines of sight are optically thin to radiation and the
central SMBH is not accreting significantly, so the radiation field
within the galaxy is nearly isotropic and produced by a spatially
distributed source (the stars themselves).  However, most galaxies
undergo brief periods of intense SMBH accretion, during which time the
radiation field is dominated by the central point source and is {\em
  not} isotropic.  During these times, most lines of sight remain
optically thin, but a few lines of sight pass through the central
dusty torus which is highly optically thick to UV/optical photons.
Furthermore, many galaxies probably have spent time in a LIRG/ULIRG state
involving intense bursts of star formation with a nearly spherical,
highly optically thick ISM in the central regions.  In this case, the
radiation field may be sourced primarily by the central point source
or the distributed stars, and in either case repeated scattering and
absorption of UV and optical photons make the radiation field tend
toward isotropy.  Overall, the galaxy does not spend much time in this
highly optically thick state, but interesting and important things are
happening during that time.  

It is crucial to note here that all of the energy absorbed in the
UV/optical bands must be re-emitted as long-wavelength photons, because
to neglect the re-emission is to ignore energy and momentum
conservation.  In particular, ULIRGs are optically thick even to
electron scattering over essentially all lines of sight, so the
diffusion of the reprocessed IR photons can have a significant impact
on the state of the gas in the galaxy \citep{murray:05, thompson:05}.

In the optically thin case, the radiation field is highly directed in
the region where the point source dominates.  This region always
exists at sufficiently small radii for a true point source.  If and
when the stellar light starts to dominate the radiation field, it
becomes nearly isotropic for radii within the region that is
producing the bulk of the photons.  For significantly larger radii,
the radiation field again becomes directed as the photons emerge from
the galaxy on nearly radial paths heading toward infinity.

If the ISM is optically thick to scattering, the radiation field
quickly becomes isotropic even if the photons come from a central
point source.  In the UV and optical part of the spectrum, the
opacities to scattering and absorption due to dust are of the same
order of magnitude.  Typically the scattering opacity is less than the
absorption opacity \citep{draine:03}, in which case the primary effect
of scattering is to make the radiation field tend toward isotropy,
rather than trapping photons so that they diffuse out of the galaxy.
Therefore in adopting a closure relation we will assume that the
radiation field becomes nearly isotropic (although still maintaining a
net outward flux) when the optical depth to scattering {\em or}
absorption is greater than unity.  If the scattering opacity is {\em
  much smaller} than the absorption opacity, this is incorrect: a
highly directed radiation field from a point source is {\em always}
highly directed even after many (absorption) optical depths.  However,
for realistic dust properties the scattering and absorption opacities
are of the same order of magnitude.

If a UV or optical photon is absorbed rather than scattered, then the
energy is reprocessed into the IR, but the energy is lost to the UV.
Thus absorption will not tend to change the angular character of the
radiation field (directed versus nearly isotropic).  However, for
declining density distributions, the column density $\int \rho \kappa
dr$ is dominated by what is occurring at small radii, whereas the
energy injected by stars, $\int \epsilon 4 \pi r^2 \, dr$ (where
$\epsilon$ is the emissivity for stellar photons), is dominated by what
happens at large radii unless the stellar distribution falls off
as $r^{-3}$ or faster.  Therefore, even if absorption dominates over
scattering (as it does in the UV), if the ISM is
optically thick, the central point source will quickly be diminished
compared to the radiation produced by stars.

Radiative transfer is difficult in general, in part because of the
high dimensionality of the problem.  The quantities of interest are
functions of spatial position, photon propagation direction,
frequency, and time, giving seven dimensions in total.  Furthermore,
the solution is subject to non-local effects: photons tend to leave a
system via optically thin ``windows'' if they are available.  A nearly
spherical gas cloud with optical depth $\opticaldepth$ along most
lines of sight and a few clear lines of sight comprising solid angle
$\Omega$ will trap photons roughly as though it had effective optical
depth $\opticaldepth_{\rm eff} = \min(\opticaldepth, \Omega/4\pi)$.
That is, in order to effectively trap photons within an optically
thick cloud, it is necessary to have a covering fraction of order
$1/\opticaldepth$ or else the photons will leak out of optically thin
``holes.''

\subsection{Mathematical treatment}
\label{sec:mathematical-treatment}
Using the physical insight provided by the discussion in 
Section \ref{sec:astrophysical-preliminaries},
we now seek a mathematical formulation of the problem
that is not overwhelmingly computationally intensive and is correct
(or nearly so) in each of the identified limits (optically thin/thick,
radiation field nearly isotropic/highly directed, dominated by
distributed source/point source, and gas distribution nearly
spherical/highly non-spherical).

We must distinguish the variables specifying the photon propagation
direction from spatial position; to this end we stipulate that
$\thetapos$ and $\phipos$ refer to spatial position while
$\thetaprop$ and $\phiprop$ refer to photon propagation direction.  Without
loss of generality, in the case of isotropic scattering, as we consider
in our treatment, the frequency-integrated equation of radiative
transfer is
\begin{equation}
  \frac{dI(\mu,\phiprop)}{ds} = - \rho (\kappa_a + \kappa_s) I + \rho \kappa_s \mi
  + \epsilon \, , 
\label{eq:rxfer-general}
\end{equation}
where $s$ is the distance along the arbitrary direction of the ray,
$\mu=\cos\thetaprop$, $\thetaprop$ is the angle between the ray propagation
direction and the $z$ axis, 
$\phiprop$ is the azimuthal angle, $I \equiv
dE/dt \, d\Omega \, dA$ is the specific intensity, $\mi \equiv \int I
\, d\mu \, d\phiprop/4\pi$ is the mean intensity, $\rho$ is the gas
density, $\kappa_a$ is the cross section for absorption per unit mass,
$\kappa_s$ is the cross section for scattering per unit mass, and
finally $\epsilon$ is the emissivity per unit solid angle
\citep{chandrasekhar:60}.  All quantities are functions of spatial
position and time, suppressed above for brevity.  

If we further adopt spherical symmetry, i.e., all quantities are
independent of spatial position $\thetapos$ and $\phipos$,
then \eqnref{eq:rxfer-general} becomes 
\begin{equation}
\mu \frac{\partial I}{\partial r} + 
\frac{1-\mu^2}{r} \frac{\partial I}{\partial \mu} = 
- \rho (\kappa_a + \kappa_s) I + \kappa_s \mi + \epsilon \, . 
\label{eq:rxfer}
\end{equation}

Here we seek a solution to this equation by taking moments in $\mu$
and integrating away the dependence of the photon field on propagation
direction.  The final ingredient is a physically reasonable closure
relation to terminate the set of moment equations, to be discussed
below.

Note that although we write the radiative transfer equation in
spherical form, the numerical method we outline does {\em not} require
spherical symmetry.  The gas density, opacities, and photon field may
vary as a function of $\thetapos$ and $\phipos$.  Further details including
the method by which this is accomplished as well as advantages and
disadvantages of our formulation are discussed in Section
\ref{sec:ray-by-ray}.  

The considerations in Section \ref{sec:astrophysical-preliminaries} 
can be summarized by the {\it ansatz} that the specific intensity to
be composed of three terms: one isotropic, one mildly anisotropic, and
one highly directed:
\begin{equation}
  I = A + B\mu + D \delta (\mu-1) \, , 
  \label{eq:intensity}
\end{equation}
where $\delta$ is the Dirac delta function and the three coefficients
(which are functions of time and radial coordinate) are adjusted as
described below.  Incorporating the last term (with $D > 0$) permits us to
allow for phases when the central point-like quasar dominates the
radiation field and to correctly treat the photon field far outside of
the galaxy where the specific intensity resembles that of a point
source.  Note that if we were to take $D=0$ in \eqnref{eq:intensity},
then we would arrive at the Eddington approximation, which is commonly
adopted in stellar interiors.  Taking \eqnref{eq:intensity} as a
generalization of the Eddington approximation allows us to interpolate
smoothly between the limits of a central point source and an optically
thick situation.

Taking moments in $\mu$ of the specific intensity to remove the
explicit dependence on the ray propagation direction gives:
\begin{equation}
  J \equiv \frac{1}{4\pi} \int_0^{2\pi} d\phiprop \int_{-1}^1  
  I(\mu,\phiprop) \, d\mu = A + \frac{D}{4}  \, , 
\end{equation}
\begin{equation}
\flux \equiv  
\int_0^{2\pi} d\phiprop \int_{-1}^1 \mu \, I(\mu,\phiprop) \, d\mu 
= \frac{4 \pi B}{3} + \pi D \, , 
\end{equation}
and 
\begin{equation}
  P \equiv \frac{1}{c} \int_0^{2\pi} d\phiprop \int_{-1}^1 \mu^2 I(\mu,\phiprop) \, d\mu 
 = \frac{4\pi A}{3 c} + \frac{\pi D}{c}  \, ,
\end{equation}
where $J$ is the mean intensity, $\flux$ is the energy flux, and $P$ is
the radiation pressure.  

We must find some physically plausible
way to decide whether the radiation field is nearly isotropic (in
which case the standard Eddington closure relation $P=u/3$ applies,
where $P$ is the radiation pressure and $u$ is the energy density) or
highly directed (in which case the flux is simply related to the
zeroth and second moments of the specific intensity).

To find the appropriate equations, we take two moments of \eqnref{eq:rxfer} 
in two different cases.  First, when the radiation field
is mildly anisotropic ($D=0$) with the closure relation $P=4\pi J/3
c$, yielding the differential equations for the photon field in the
classic Eddington approximation:
\begin{equation}
  \frac{dL}{dr} = 4 \pi r^2 (\dot{E} - 4 \pi \rho \kappa_a \mi) \, , 
\end{equation}
\begin{equation}
  \frac{d\mi}{dr} = - \frac{3 \rho (\kappa_a+\kappa_s) L}{16 \pi^2 r^2} \, .
  \label{inten-eddington}
\end{equation}
Second, when the radiation field is highly directed ($A=B=0, D=1$)
with the closure relation $P=4 \pi J/c $, appropriate for a point
source.  In this case the equation for $dL/dr$ is the same and the
equation for mean intensity becomes
\begin{equation}
  \frac{d\mi}{dr} = - \frac{2 \mi}{r} 
- \frac{\rho (\kappa_a+\kappa_s) L}{16 \pi^2 r^2} \, .
\label{inten-point-source}
\end{equation}
The first term on the right-hand side gives the expected $r^{-2}$
fall-off of the mean intensity if the radiation field is primarily
directed (e.g. outside of the galaxy).

We combine equations \eqntworef{inten-eddington}{inten-point-source}
for $dJ/dr$ by introducing the variable $t$ that interpolates between the
optically thin (radiation field like a point source) and optically
thick (radiation field nearly isotropic) cases.  This gives:
\begin{equation}
  \frac{dL}{dr} = 4 \pi r^2 (\dot{E} - 4 \pi \rho \kappa_a \mi) \, , 
\label{eq:lum-rxfer}
\end{equation}
\begin{equation}
  \frac{d\mi}{dr} = - \frac{2 \mi t}{r} 
- \frac{(3-2t) \rho (\kappa_a+\kappa_s) L}{16 \pi^2 r^2} \, , 
\label{eq:inten}
\end{equation}
where $t=0$ when the radiation field is nearly isotropic and $t=1$
when it is highly directed.  In practice, we take $t$ as a
function of position to be:
\begin{equation}
  1-t(r) \equiv \frac{1} {1 + e^{5 (1-\opticaldepth_{\rm out})} + 
    e^{5(1-\opticaldepth_{\rm in})}} \, , 
\label{eq:thin-thick}
\end{equation}
where 
\begin{equation}
  \opticaldepth_{\rm in}(r) = \int_0^r \rho (\kappa_a + \kappa_a) \, dr \, ,
\end{equation}
and 
\begin{equation}
  \opticaldepth_{\rm out}(r) = \int_r^\infty \rho (\kappa_a + \kappa_a) \, dr
  \, .
\end{equation}
Evaluating $t$ as $r$ becomes very large or very small gives $t=1$ at
the inner boundary (near the AGN) and at the outer boundary (far from
the galaxy).  If the ISM is optically thick there will be some region
in between where $t$ is nearly 0.  This choice expresses our
expectation about where the radiation field should be primarily
directed versus primarily isotropic.  The factors of five in
\eqnref{eq:thin-thick} are chosen to ensure that the value of $t$
switches from the optically thin limit to the optically thick limit
fairly quickly near optical depths of unity.  The transition
effectively occurs between optical depths of 0.8 and 1.2.

Finally we specify the boundary conditions.  As in the study of
stellar interiors, our second order differential equation for $L(r)$
requires two boundary conditions, one at the center and one at the
outer surface.  Near the center of the
simulation, the outgoing luminosity is only that provided by the
central point source, the AGN.  Far from the galaxy, the radiation
field is expected to again look like a point source as
scattering/absorption become negligible and all of the photons leave
the galaxy.  The two natural boundary conditions are then
\begin{equation}
  L(r_{\rm min}) = L_{\rm BH} \, , 
  \label{eq:bc1}
\end{equation}
and
\begin{equation}
  J(r_{\rm max}) = \frac{L(r_{\rm max})}{16 \pi^2 r_{\rm max}^2} \, .
  \label{eq:bc2}
\end{equation}

The force on the gas in a computational cell is
\begin{equation}
  \Delta \force = \frac{L(r) d\opticaldepth dA}{4 \pi r^2 c} \, ,
  \label{eq:forces}
\end{equation}
where $d\opticaldepth$ is the optical depth (scattering plus
absorption) of a given cell and $dA$ is the area of the cell
perpendicular to the radiation flux vector.

\subsubsection{Comments on the spherical symmetry assumption}
\label{sec:ray-by-ray}
We are primarily interested in what happens to the ISM as a function
of radius rather than angle.  Angular effects are important and it is
of crucial importance to break spherical symmetry so that hydrodynamic
instabilities such as convection and the Rayleigh-Taylor instability
can operate.  However, we are concerned mainly with {\em whether} gas
makes it to small radii (and accretes onto the black hole) or to large
radius (and escapes the galaxy), not {\em where} on the sphere the gas
enters the SMBH or leave the galaxy.  Thus we neglect all angular
radiative forces and formulate the radiative transfer as occurring
exclusively along rays from the center of the galaxy.

Similar approximations have been used in numerical simulations of photon
and neutrino transport in supernovae for some time \citep[e.g.][and
references therein]{brandt:11}.  One approach is to spherically
average the matter distribution, solve for the radiation field, and
then calculate forces on matter via \eqnref{eq:forces}.  Another
approach is to treat each ray as a separate problem, in which case all
intensive quantities (gas density, emissivity of stellar photons, etc)
are treated as though they obtained over the whole sphere.

We found that the first approach has dramatic failure modes where it
wildly overestimates the outward radial forces on the gas in certain
situations.  As an example, consider an optically thick cloud covering
a small solid angle in an otherwise optically thin medium.  In
reality, UV photons deposit momentum in a thin layer at the surface of
the cloud, and the total momentum imparted to the cloud depends on the
solid angle of the cloud.  However, spherically averaging the matter
distribution may result in the ISM becoming optically thin at the
cloud radius, and the solution for the photon field will indicate that
the UV flux is high throughout the cloud.  In this case, essentially
all of the dust grains in the cloud are able to absorb UV photons and
the momentum imparted to the cloud will be wildly overestimated.

Instead, treating each ray as a completely independent problem has
much better behavior in the sense that when the algorithm makes an
error in computing the force on a particular gas cell, fractional
error tends to be of order unity rather than orders of magnitude.  In
the previous example, this method gives the correct force exerted by
UV photons.  For the IR photons, the method assumes that they must
traverse the cloud on their way out of the galaxy, while in reality
they would leave the system via the optically thin ambient medium.
Therefore the method makes a a fractional error of order the IR
optical depth when computing the force.  

One drawback of the latter approach is that if a particular ray (say,
near the pole) has deficient star formation compared to the rest of
the galaxy, then the forces exerted by the stellar photons will be
anemic compared to the true solution, where stars from near the
midplane emit photons that help push on gas near the poles.
Therefore, we average source terms over a sphere before finding the
solution to the radiative transfer equations on each ray.

Note that this method allows the radiation field to differ in an
unlimited way from one ray to the next.  This is expected for example
in the UV where the scattering opacity is less than the absorption
opacity, and the interior of an individual ``cloud'' along a line of
sight can be shielded from the AGN UV luminosity.  However, in the IR
we consider the absorption opacity to be zero, and it no longer makes
sense that the radiation field could differ substantially on adjacent
rays.  Therefore we also spherically average the IR source terms (in
particular IR generated by absorption of UV or optical) before solving
for the luminosity as a function of radius in the IR.

\subsection{Numerical implementation}
\label{sec:implementation}
The numerical solution of the radiative transfer equation is
complicated by the fact that one boundary condition is specified at
$r_{\rm min}$ while the other is specified at $r_{\rm max}$ (note
\eqntworef{eq:bc1}{eq:bc2}).  Thus, integration from either side
requires a guess that must be refined.  This is
identical to the situation in treating the radiative transfer in
stellar interiors.  Conceptually, this is simply because we know the
central luminosity, but we do know know the {\em intensity} of the
radiation field near the center.  If the scattering opacity is large,
photons will build up in the galaxy forming a nearly isotropic
radiation field until the gradient of the photon density is sufficient
to drive photon diffusion out of the galaxy at the rate required to
carry the specified luminosity.

In fact, the numerical solution to the radiative transfer equation in
stellar interiors is considerably simpler than in the present problem
because within stars, one can exploit the fact that the plasma is
everywhere close to local thermodynamic equilibrium.  The plasma is
assumed to quickly thermalize any energy sources or sinks so that the
plasma is well-described by a single temperature at any point.  In
galaxies, the photon field is not close the thermodynamic equilibrium
and we must track energy as it is transferred from x-rays to UV
photons and eventually to IR photons.  Treating stellar interiors in
an analogous way would mean tracking each energy band of photons
separately as the star's luminosity was transported through the
plasma, from the primary MeV photons produced by fusion reactions, to
the UV photons that eventually emerge from the photosphere.

To find the solution of \eqntworef{eq:lum-rxfer}{eq:inten} subject to the
correct boundary conditions \eqntworef{eq:bc1}{eq:bc2}, the ``shooting
method'' with Newton-Raphson iteration \citep{press:92} works well up
to optical depths of a few.  Beyond that, the absorbing gas acts as a
very effective ``screen'' that hides conditions near the center of the
simulation from those at the outer edge.  This makes it very difficult
to find the correct solution matching both boundary conditions at
once.  We would like to simulate AGN feedback in galaxies with gas
surface density up to Compton-thick, implying an optical depth to dust
absorption in the UV of several thousand.  The shooting method is too
unstable to use in this case.  

It is true that given such a large optical depth, very little of the
primary UV radiation escapes the galaxy; it comes out in the IR.
However, it is incorrect to discard the primary UV radiation
completely, or to immediately transfer it to the IR.  While this would
correctly give the emergent energy, the UV radiation can have a very
large effect near the center of the galaxy before it is transferred
to the IR.  For for solar composition and dust/gas ratios, they can
limit the accretion rate to $3\times 10^{-4} L_{\rm Edd}$.

At large optical depths, the more powerful ``relaxation'' iterative
method \citep{press:92} is required to adjust the solution over the
entire domain of integration while ensuring that the proper boundary
conditions are met at both ends of the interval.  Newton's method is
used to iterate over trial solutions to arrive at the solution to the
differential equations, where the independent variables of the Newton
iteration are taken to be the values of $L$ and $\mi$ at each spatial
point.  The method takes the coupling between variables at neighboring
points into account, allowing the entire solution to be updated
without overcompensating (which the shooting method is prone to do).
Given a good initial guess for the iteration, the relaxation method
finds a reasonably accurate solution even for very large optical
depths.  The initial guess is conveniently provided by the solution
that pertained during the previous simulation time step.

Two additional numerical schemes are defined in appendicies.  The
method described in Appendix \ref{sec:complex-rxfer} allows one to
avoid using the variable $t$ and its somewhat \textit{ad hoc}
definition in \eqnref{eq:thin-thick} at a cost of one additional
additional differential equation to solve and a somewhat more opaque
relationship between the equations and their physical content.  The
algorithm described in Appendix \ref{sec:simplified-rxfer} is an
attempt to preserve the basic physics of the method while transforming
the system of differential equations from a boundary value problem to
an initial value problem.  This allows one to use a substantially
simpler numerical method to obtain the solution.

\section[]{Hydrodynamical Simulations}
\label {sec:hydro-simulations}
For the full simulations, we will use the 2D numerical code described
in \citet{novak:11}.  The code implements sources and
sinks of mass, energy and momentum relevant for an early-type galaxy
including stellar mass loss from planetary nebulae, Type 1a and Type
II supernovae.  A full description of the input physics can be found
in \citet{novak:11} and \citet{ciotti:12}.  We describe the simple
treatment of dust creation and destruction used in the present work.
We refer readers to \citet{hensley:12-preprint} for a more complete
implementation of dust physics applied to one dimensional models.  

\subsection{Timescales}

The light crossing time of the galaxy is by far the shortest timescale
in the system and therefore if the ISM is optically thin, photons are
considered to act instantaneously for the purpose of our simulation.
When the ISM becomes optically thick, the radiative diffusion time is 
\begin{equation}
  \timescale_{\rm diff} \simeq \frac{R \opticaldepth}{c} \, ,
\end{equation}
where $\opticaldepth$ is the optical depth.  Typically 
the temperature of the ISM is not far from the virial temperature of
the galaxy, so that the dynamical time and the sound crossing
time are nearly equal.  The latter is
\begin{equation}
  \timescale_{\rm S} \simeq \frac{R}{c_s} \, .
\end{equation}
Thus the radiative diffusion time will be shorter than the dynamical
time and sound crossing time as long as $ \opticaldepth \le c/c_s \simeq
1000$.  For our purpose radiative diffusion takes place primarily in
the IR, and the maximum optical depth in the IR is of order 10.
Therefore the radiation field is considered to reach equilibrium
instantaneously, and radiative forces are considered to act
instantaneously.   That is to say that we solve all of the
radiation transfer equations assuming a steady state.

\subsection{Photon energy bands}

We divide the radiative energy output of the AGN into broad bands where the relevant
physics is significantly different: IR (below 1 eV), Optical (for our
purposes, non-ionizing: 1-13 eV), UV including soft X-rays where the
absorption cross-section is greater than the Thompson cross-section
(13-2000 eV), and hard X-rays where the interactions are dominated by
Compton scattering ($\ge$ 2 keV).  The assumed division of the AGN
output into the different bands is given in Table \ref{tab:bands}.

\begin{table}
  \centering
 \begin{minipage}{\columnwidth}
   \caption{Assumed unabsorbed QSO energy output by band
     \citep{sazonov:04} and dust opacities for solar metallicity gas with
     Milky Way dust-to-gas ratio expressed in terms of the electron
     scattering opacity, $\kappa_{\rm ES} = 0.4 \unit{cm}^2 \unit{g}^{-1}$}
   \label{tab:bands}
  \begin{tabular}{@{}llll@{}}
    \hline
    Band    &  Energy & Fraction & Dust Opacity \\ 
    \hline
    X-ray   & E \textgreater 2 keV & 0.1 \\
    UV      & 13 eV \textless E \textless 2 keV & 0.35 & 3430 $\kappa_{\rm ES}$ \\ 
    Optical & 1 \textless E \textless 13 eV & 0.25 & 857 $\kappa_{\rm ES}$ \\
    IR      & E \textless 1 eV & 0.3 & 5.71 $\kappa_{\rm ES}$ \\
    \hline
  \end{tabular}
\end{minipage}
\end{table}

The radiative transport in the X-ray band is relatively simple.  Few
AGN are observed to be Compton thick and our simulations
rarely reach such a high column density, so the ISM is always treated
as optically thin.  X-rays dominate the heating of the gas to
temperatures of order the galactic virial temperature.  We treat
heating and cooling due to photoionization and Compton heating/cooling
by atomic lines via the formulae given in \citet{sazonov:05}.  X-rays
also affect the gas momentum via Compton scattering and
photoionization.  Compton scattering by X-rays is handled along with
the bolometric AGN luminosity owing to the wavelength independence of
the Compton cross section.  Momentum transferred to the ISM by
photoionization is taken to be
\begin{equation}
  \frac{dp}{dt} = \frac{H}{c} \, ,
  \label{eq:photoionization-force}
\end{equation}
where $H$ is the photoionization heating rate.  It should be noted
that this is a lower limit to the total effect of photoionization on
the momentum of the ISM.  The Sazonov formulae give the photo heating
rate due to X-rays, but neglect the heating at lower energies due,
e.g. to optical and UV photons.  

An exact treatment of the effect of optical and UV photons
necessitates tracking the temperature as well as the ionization state
of the gas---this quickly becomes prohibitively complex.  Instead we
note that X-rays are required to raise the gas temperature to the same
order as the virial temperature for a massive galaxy, and therefore
drive an outflow into the IGM.  Lower energy photons may affect the
detailed state of the gas in the galaxy, but they cannot affect
whether or not the gas remains bound to the galaxy.  Compton heating
can be important if the Compton temperature of a given object is
large.  However, for typical objects Compton heating and cooling are
sub-dominant.  While optical and UV photons acting on atomic resonance
lines are unable to raise the gas temperature to the galactic virial
temperature, they nevertheless contribute to the force on the ISM.
For the time being we neglect them.

Radiative transport in the IR is similarly simple.  At long
wavelengths, the dust absorption opacity is much greater than the
scattering opacity \citep{draine:03}.  However, the absorbed energy is
re-radiated by the dust grains and thus the process is more similar to
scattering than true absorption.  In order to conserve energy, we
simply treat the process as scattering.  Since no energy is ever lost
from the IR band and must leave the galaxy eventually, the IR luminosity
as a function of radius is simply the integral of all source terms
within the radius.

The UV and optical bands are considerably more complex.  Each band
satisfies a separate set of differential equations given by
\eqntworef{eq:lum-rxfer}{eq:inten}.  The boundary conditions and
method of solution are as described in
Sections \ref{sec:mathematical-treatment}
and \ref{sec:implementation}, applied to each band separately.
Radiation absorbed in both bands contributes to the IR.

\subsection{Opacities}

At a minimum, photons emitted from the stars and AGN impart momentum
to the ISM via Compton scattering.  The Compton cross section is
wavelength independent and coherent up to energies of order 511 keV,
making the effect easy to treat numerically.  The force on a cell is
given by \eqnref{eq:forces} where $L$ is the bolometric luminosity of
stars and the AGN within the radius $r$.  This is correct even when
the ISM is Compton thick because the process involves only scattering and
no absorption (up to energies approaching 511 keV).

The effect of absorption by dust is more complicated because the
opacities depend strongly on wavelength, and hence it is necessary to
track the luminosity as a function of photon energy and radius.  
The dust opacities for each band for solar metallicity metallicity gas
with a Milky Way dust-to-gas ratio are given in Table \ref{tab:bands}.  

Finally, outgoing photons can be absorbed by resonant atomic lines.
In the case that the atom enters an excited state, the photon produced
by the decay of the excited state is emitted isotropically.  Thus the
photons impart momentum, but no net energy to the gas.  These
bound-bound transitions can have very large opacities, but only over a
narrow range of frequencies, so we neglect this effect in the present
work.  In the case of photoionization, the energy in excess of the
binding energy goes into heating the gas, imparting both energy and
momentum to the gas.  For optical and UV photons, accurate calculation
of the photoionization opacity requires detailed knowledge of the
temperature and ionization state of ISM gas that the present
simulations lack.  Therefore we leave the inclusion of photoionization
opacity in the optical and UV bands for future work.  For the
high-energy photons that dominate the heating of the gas \citep[see
the discussion in the previous section and][]{sazonov:04}, the
photoionization force is included through
\eqnref{eq:photoionization-force}, as discussed above.

\subsection{Dust creation and destruction}
Dust in ETGs is thought to be produced primarily in planetary nebulae
and asymptotic giant branch outflows, and destroyed in hot gas by
sputtering.  \citet{draine:79} give a theoretical estimate of the
dust destruction time as a function of grain size:
\begin{equation}
\timescale_{\rm destroy} \equiv  \frac{a}{|\dot{a}|} = 10^5 \unit{yr} \, 
a_{0.1} n_1^{-1} (1 + T_6^{-3}) \, ,
\label{eq:dust-destruction}
\end{equation}
where $a_{0.1} = a/0.1\unit{$\mu$m}$, $n_1 = \rho/m_p \unit{cm}^{-3}$, and
$T_6 = T/10^6 \unit{K}$.  Dust is also commonly observed to be
produced via supernovae.  Therefore, when gas is added to the
computational grid by explicit stellar source terms, we also add dust
in proportion.

\citet{draine:09} has recently argued that dust grain growth is
predominantly by gas-phase growth in the ISM rather than in
supernovae.  If this is the case, we should add an explicit dust
source term that causes the dust to gas ratio to relax to a specified
value when the gas is cold and star forming, regardless of whether all
of the gas in the cell is actually processed through supernovae.  

In this scenario, dust grains are thought to be built up by accretion
of single atoms from the gas phase, the inverse of the process that
destroys grains at high temperature.  The only difference is that
grain growth requires that a dust grain collide with a metal atom
rather than a more abundant hydrogen atom.  Metal atoms are both less
abundant and move more slowly in thermal equilibrium.  Therefore the
timescale for creation is easily obtained by scaling from the
destruction timescale:
\begin{equation}
  \timescale_{\rm create} = 10^5 \unit{yr} \, A^{3/2} Z^{-1}
a_{0.1} n_1^{-1} T_6^{1/2} (1 + 10 \, T_6)  \, ,
\end{equation}
where $A$ is the atomic weight of the atoms in question (here taken to
be $A=12$ for Carbon), $Z=\rho_{\rm metal}/\rho$ is the metallicity of
the gas, and the $(1+10\,T_6)$ term is added to ensure that dust
creation takes place only at temperatures below $10^{5}$ K in the
simulation.

The dust density in a given cell is represented by a passively
advected tracer field with appropriate source terms for dust creation
and destruction.  The source and sink terms for dust in a given cell
are
\begin{equation}
\dot{\rho}_{\rm dust} = 
- \frac{\rho_{\rm dust}}{\timescale_{\rm destroy}} + 
\frac{\rho Z-\rho_{\rm dust}}{\timescale_{\rm create}} \, .
\end{equation}
Dust grain growth terminates when all of the metals are in dust
grains.  In this paper the treatment of dust is significantly improved
with respect to the basic approach in the series of papers by
\citeauthor{ciotti:01}; however, in \citet{hensley:12-preprint}, the
dust treatment is discussed in full generality, albeit in the context
of 1D simulations.

Including advection, the partial differential equation for the dust
density is
\begin{equation}
\frac{\partial \rho_{\rm dust}}{\partial t}
+ \frac{\partial (v_r \rho_{\rm dust})}{\partial r}
+ \frac{1}{r} \frac{\partial (v_\theta \rho_{\rm dust})}{\partial \theta}
   = - \frac{\rho_{\rm dust}}{\timescale_{\rm destroy}} + 
\frac{\rho Z-\rho_{\rm dust}}{\timescale_{\rm create}} \, .
\end{equation}

The radiative force on a cell due to absorption by dust is obtained
from \eqnref{eq:forces}
\begin{equation}
  \Delta \flux = \frac{\rho_D}{\rho} \frac{L(r) d\opticaldepth dA}{4 \pi r^2 c} \, ,
\end{equation}
where $d\opticaldepth$ is the optical depth {\em if} there were no
dust depletion, and the effect of dust depletion is taken into account
by the density ratio.

One possible improvement to our treatment of dust creation and
destruction would be to track the dust size distribution directly via
moments.  Our current implementation essentially assumes that all dust
grains are maintained with a given fiducial size of 0.1 \unit{$\mu$m} and
are then eroded via sputtering or grown via metal atom deposition.  

\subsection{Stellar source terms}

\begin{figure*}
  \includegraphics[width=1.3\textwidth]{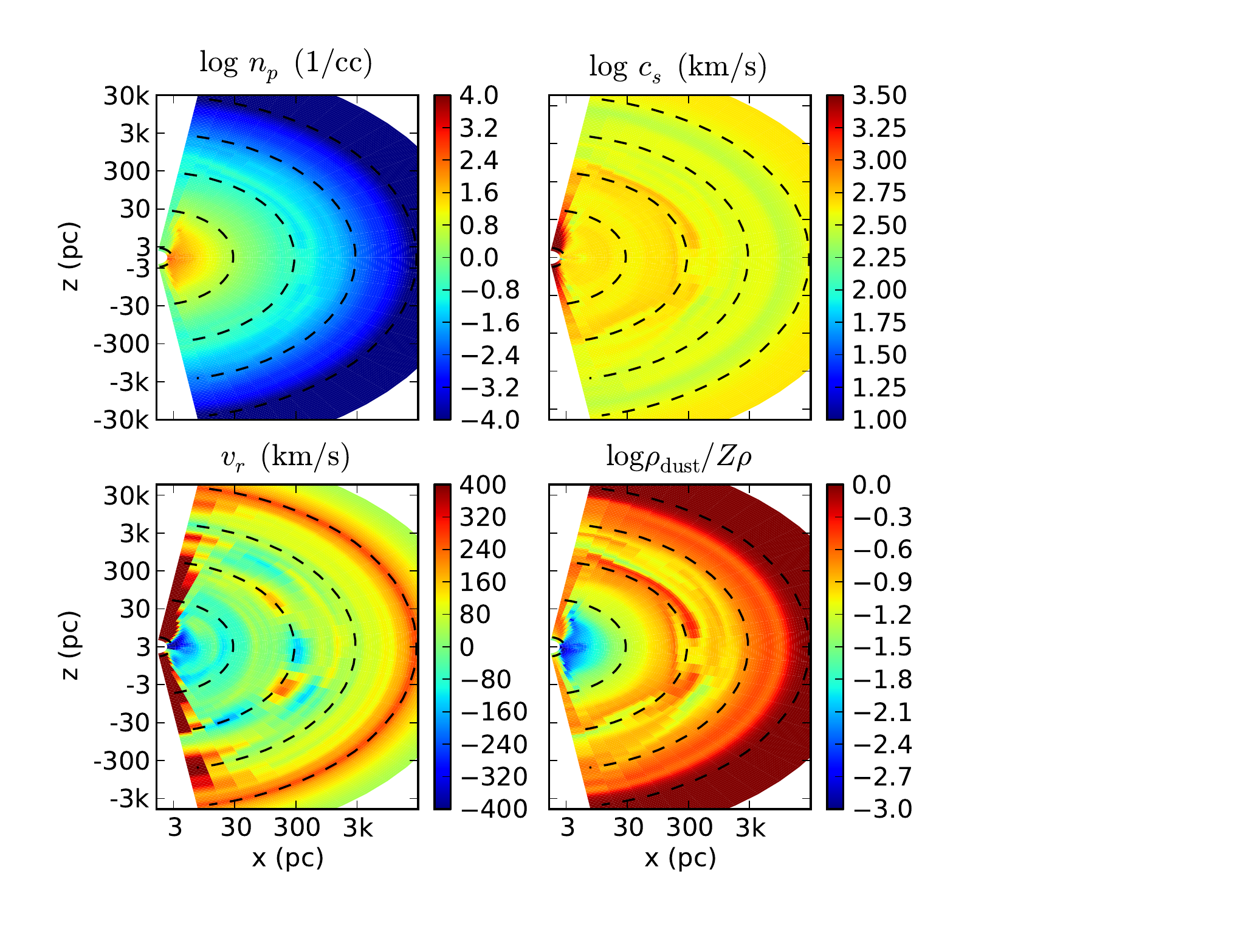}
  \caption{A snapshot of the A2 simulation during a quiescent phase.
    Upper left, log gas density in protons per cubic centimeter.
    Upper right, log sound speed.  Lower left, radial velocity.  Lower
    right, the log of the dust to gas ratio normalized to the value in
    the Milky Way.  Gas that enters the ISM via the broad-line wind
    from the AGN is assumed to be dust rich.  At small radii,
    densities and temperatures are large enough to efficiently destroy
    the dust grains, resulting in dust suppression by a factor of up to
    1000.}
\label{fig:snap1}
\end{figure*}

\begin{figure*}
  \includegraphics[width=1.3\textwidth]{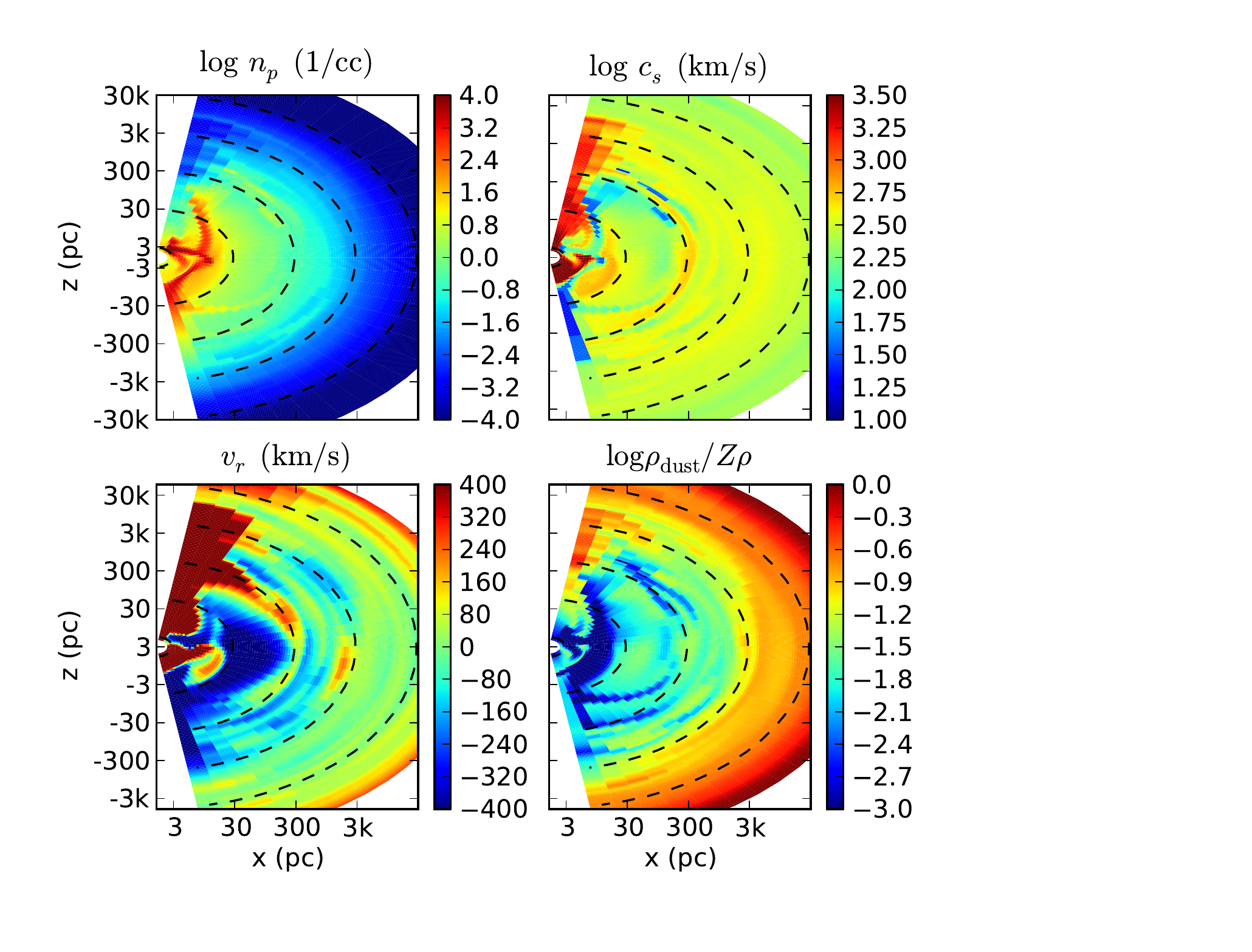}
 \caption{A snapshot of the A2 simulation during a black hole
   outburst.  All gas near the center of the galaxy is depleted of
   dust by a factor of at least 30 and up to 1000.}
\label{fig:snap2}
\end{figure*}

Since stars are sources of UV and optical light that contribute to the
solution to the radiative transfer equation discussed here, we briefly
review our implementation of star formation and generation of star
light.  Gas that forms stars also contributes to the energy balance
via supernovae, our implementation of which is discussed in detail in
\citetalias{ciotti:07}.

Gas is removed from the galaxy by star formation at the instantaneous
rate given by the standard formula based on the
\citet{kennicutt:98} law:

\begin{equation}
  \dot{\rho}_{\rm gas} = - \frac{\eta_{\rm form} \rho
  }{\timescale_{\rm form}} \, , \qquad \eta_{\rm form} = 0.1 \, , 
\end{equation}
\begin{equation}
  \timescale_{\rm form} = \max(\timescale_{\rm cool}, \timescale_{\rm dyn}) \, ,
\end{equation}
\begin{equation}
  \timescale_{\rm cool} = \frac{3 k_B T}{2 n \Lambda(T)} \, , \qquad \timescale_{\rm dyn} = \min(\timescale_{\rm jeans}, \timescale_{\rm rot}) \, ,
\end{equation}
\begin{equation}
  \timescale_{\rm jeans} = \sqrt{\frac{3 \pi}{32 G \rho}} \, , \qquad \timescale_{\rm rot} = \frac{2 \pi r}{v_c(r)} \, ,
\end{equation}
where $n^2 \Lambda(T)$ is the volumetric cooling function and $v_c(r)$
is the circular velocity at radius $r$ in the galaxy
(\citet{ciotti:12}).

Mass that is transferred from gas to stars at radius $r$ is first
transferred to a ``forming'' state by convolving the star formation
rate with an exponential kernel with a time constant of $3 \unit{Myr}$,
designed to account for the finite minimum star formation given by the
Kelvin-Helmholtz timescale as well as the minimum stellar lifetime
given by the combination of the Eddington luminosity and the nuclear
energy available to a massive star:
\begin{equation}
  \dot{\rho}_{\rm forming}(t) = - \frac{1}{\timescale_{\rm KH}} \int_{0}^t \dot{\rho}_{\rm gas} (t')
  e^{-\frac{t'}{\timescale_{\rm KH}}} \, dt' \, .
\end{equation}

These stars also contribute photons to the UV/optical luminosity of
the galaxy via convolution with a similar exponential kernel:

\begin{equation}
  \frac{dL_B}{dV}(t) = - \frac{1}{\timescale_{B}}
  \int_{0}^t \epsilon_B c^2 \dot{\rho}_{\rm gas} (t')
  e^{-\frac{t'}{\timescale_B}} \, dt' \, ,
\end{equation}
where $B$ refers to band and takes the values UV or optical.  Values
for $\epsilon_B$ and $\timescale_B$ for each band are given in Table
\ref{tab:stars}.

\subsection{Radiative cooling and heating}

Photons generated by collisional excitation of atomic lines or
free-free radiation can only cool the ISM insofar as the photons can
actually escape.  If the galaxy becomes optically thick to IR
radiation, then these photons can contribute significantly to the
forces experienced by gas in the ISM as they diffuse out of the
galaxy.  Therefore we add the cooling radiation to the distributed UV
(optical) source terms if the local gas temperature is above (below)
$10^5 \unit{K}$.  The radiative formulae are discussed in
\citet{ciotti:12}.

\begin{table}
 \centering
 \begin{minipage}{\columnwidth} 
  \caption{Parameters governing stellar luminosities (see text for definitions)}
  \label{tab:stars}
  \begin{tabular}{@{}ll@{}}
    \hline
    Band    &  Fraction \\ 
    \hline
    $\epsilon_{\rm UV}$  &  $8.65\times 10^{-5}$ \\
    $\epsilon_{\rm Opt}$ &  $1.24\times 10^{-3}$ \\
    $\timescale_{\rm UV}$      &  $2.57 \unit{Myr}$ \\
    $\timescale_{\rm Opt}$     &  $154 \unit{Myr}$ \\
    \hline
  \end{tabular}
\end{minipage}
\end{table}

\subsection{Broad line wind}
Our simulations also include a treatment of the broad-line wind
generated near the SMBH.  This is implemented as an inflow term on the
edge of the computational grid \citep{novak:11, ciotti:12}.  Although the present paper focuses on
radiative AGN feedback, we summarize our implementation of mechanical
feedback because it has a large effect on the regulation of black hole
growth.  See \citet{novak:11} and \citet{ostriker:10} for further
discussion of mechanical feedback.

The BAL originating near the SMBH provides energy, momentum, and
mass from a wind to the ISM according to the equations \citep{ostriker:10}:
\begin{equation}
  \dot{M}_{\rm BH} = \frac{\dot{M}_{\rm infall}}{1+\eta} \, , \qquad \eta = \frac{2 \epsilon_W c^2}{v_W^2} \, , 
\end{equation}
\begin{equation}
  \dot{E}_W = \epsilon_W \dot{M}_{\rm BH} c^2 \, , \qquad \dot{P}_W = \frac{2 \epsilon_W c^2  \dot{M}_{\rm BH}}{v_W} \, , 
\end{equation}
and
\begin{equation}
  \dot{M}_W = \frac{2 \epsilon_W c^2  \dot{M}_{\rm BH}}{v_W^2} \, , 
  \label{eq:mdot-w}
\end{equation}
where $v_W$ is the velocity of the BAL, taken here to be 10,000 km s$^{-1}$.  

Independently of the mechanical feedback model, the radiative
luminosity of the AGN is given by
\begin{equation}
  L = \epsilon_{\rm EM} \dot{M}_{\rm BH} c^2  \, ,
  \label{eq:lum-agn}
\end{equation}
where the electromagnetic efficiency is given by the advection
dominated accretion flow inspired
\citep{narayan:94} formula:
\begin{equation}
  \epsilon_{\rm EM} = \frac{\epsilon_0 A \dot{m}}{ 1 + A \dot{m}}
  \, ,
  \label{eq:epsilon-em}
\end{equation}
and $A=100$ and $\epsilon_0=0.1$.  The dimensionless mass accretion rate is
\begin{equation}
  \dot{m} = \frac{\dot{M}_{\rm BH}}{\dot{M}_{\rm Edd}} = 
  \frac{\epsilon_0 \dot{M}_{\rm BH} c^2}{L_{\rm Edd}} \, ,
\end{equation}
where $L_{\rm Edd}$ is the Eddington luminosity.

Although our other work has considered models where $\epsilon_W$ is a
function of the accretion rate, in the present work we assume that
$\epsilon_W$ is a constant with values given in Table
\ref{tab:epsilon}.  In previous papers these are called A models, and
we maintain that nomenclature here.  Our A0 model has the same value
of $\epsilon_W$ as that adopted by \citet{dimatteo:05}, and our A2
model has a lower value, favored by recent observations
\citep{moe:09,arav:12-preprint}.

\begin{table}
  \centering
 \begin{minipage}{\columnwidth}
  \caption{Values for the mechanical wind efficiency considered in this
  work.}
  \label{tab:epsilon}
  \begin{tabular}{@{}ll@{}}
    \hline
    Model &  $\epsilon_W$ \\ 
    \hline
    A0 & $5\times 10^{-3}$ \\
    A2 & $10^{-4}$ \\
    \hline
  \end{tabular}
\end{minipage}
\end{table}

\subsection{Galaxy model}
The total gravitational potential of the model galaxy is assumed to be
a singular isothermal sphere plus a point mass for the central BH.
This is good agreement with observations of the total mass profile of
early-type galaxies \citep{gavazzi:07, gavazzi:08}.  For simplicity we
maintain this model in to the smallest radii, although more
complicated models may be more appropriate inside of a fraction of the
half-light radius.  The velocity dispersion parameter of the
isothermal potential is 260 km s$^{-1}$.  The gas is not
self-gravitating.  The stellar distribution is given by a Jaffe
profile with a total mass of $3\times 10^{11} M_\odot$ and a projected
half-mass radius of 6.9 kpc.  The mass-to-light ratio is assumed to be
spatially constant and is equal to 5.8 in solar units in the $B$ band
at the present time.  The structural and dynamical properties of these
galaxy models are discussed in detail in \citet{ciotti:09-dynamics}.

We note that although the galaxy model is spherically symmetric and
the physics as implemented is either sperically symmetric or at least
symmetric under inversions of the $z$ axis, numerical noise is quickly
amplified chaotically in the simulations so that the physical state
does not remain up/down symmetric for very long.

\section[]{Results}
\label{sec:results} 

All of the simulations presented here use the method described in Sections
\ref{sec:mathematical-treatment} through \ref{sec:implementation} to
solve for the photon field, where the true boundary value problem is
solved via the relaxation method.  

Figures \ref{fig:snap1} and \ref{fig:snap2} show two snapshots from
the ``fiducial'' A2 simulation with mechanical AGN feedback efficiency of
$\epsilon_W=10^{-4}$ and all of the physics thus far described enabled.  The dust
to gas ratio is initialized to the Milky Way value throughout the
galaxy.  This simulation uses the ``standard'' recipe for dust grain
creation where dust is created in supernova and dust grains do not not
grow via collisions with gas-phase metal atoms in the ISM.  However,
all gas that enters the simulation grid does so with a Milky Way
dust-to-gas ratio.  

Figure \ref{fig:snap1} shows that near the center of the galaxy, dust
is efficiently destroyed so that forces on gas due to absorption of
photons by dust grains are not very effective at preventing gas from
falling into the center of the galaxy.  At large radius, the
temperatures are sufficient for sputtering to be effective, but the
gas densities are low enough that the grain destruction timescale is
long.

In this picture, cold gas is generated by unstable cooling of the ISM.
Cold gas loses pressure support and falls toward the center of the
galaxy on a free-fall timescale.  Cold gas is generated from hot gas,
and the dust grains in the hot gas have been eroded by sputtering for
for at least a cooling time.  Therefore radiative forces on dust
grains do not play a major role in the dynamics.  Even if dust grains
are replenished in the ISM without being processed through stars as
suggested by \citet{draine:09}, the grain growth timescale is at best
comparable to the infall timescale.  Given their initially depleted
state, dust grains must have many growth timescales available in order
to replenish the dust.

If radiative forces on dust are to play a role in black hole
accretion, then the gas feeding the black hole must either never have
been hot, or it must be cold for a long time before approaching
the black hole.  If the gas had been hot, it cannot return to the cold
phase until a cooling time has passed, sufficient to significantly
deplete the dust grains.  Cold gas that is not rotationally supported
falls to the center of the galaxy too quickly to build up significant
dust.  However, dust lanes are observed in a significant fraction of
elliptical galaxies \citep{hawarden:81,vdokkum:95}.  If the black hole
is fed by gas processed through a rotationally supported
galactic-scale disk, then dust grains would have a chance to build up
while the gas spends a long time orbiting the black hole.

\begin{figure}
  \includegraphics[width=0.65\textwidth]{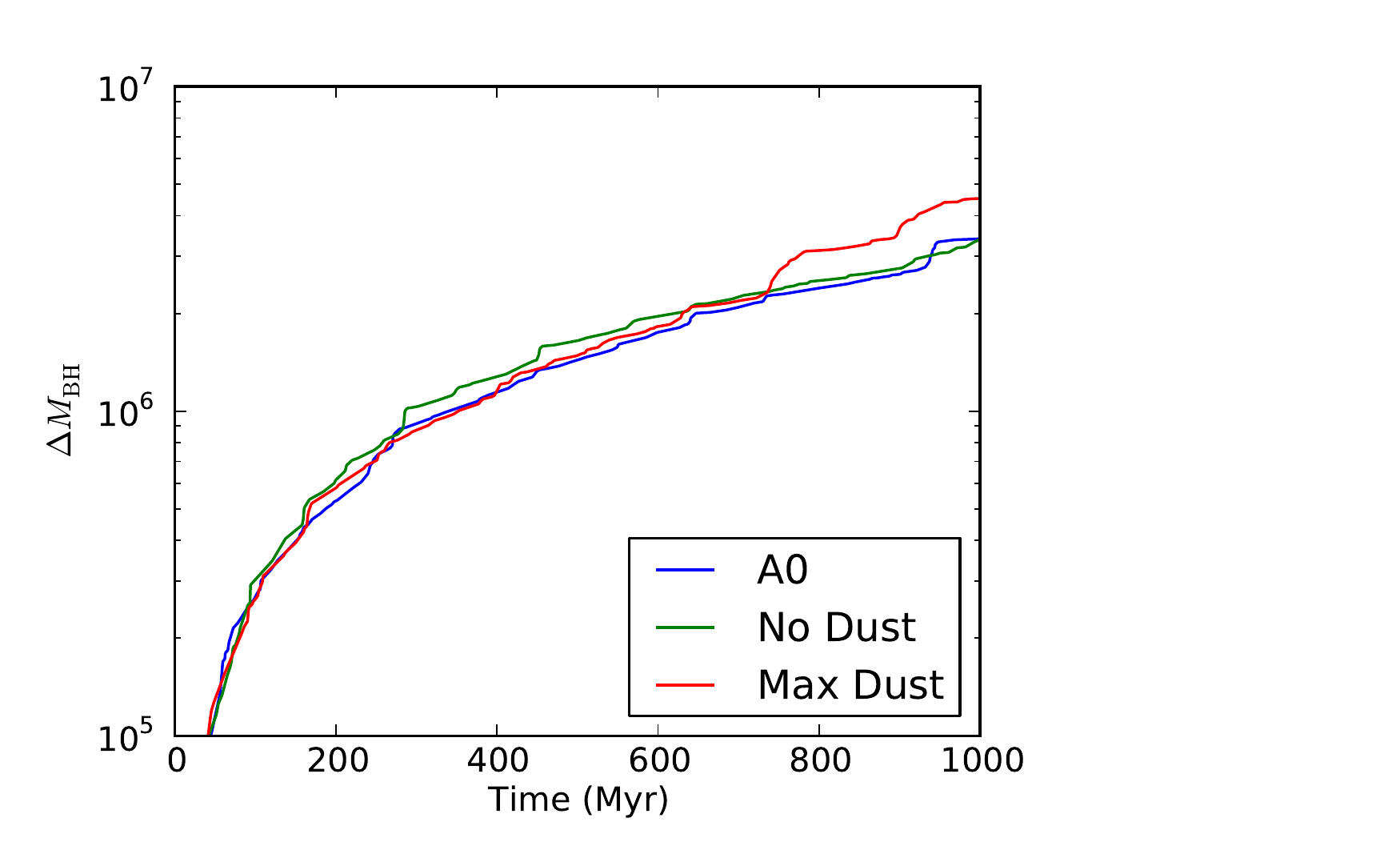}
  \caption{Black hole growth over 1 Gyr in the A0 simulation for
    different assumptions about the dust.  The blue line shows the
    result of tracking dust density with a tracer field and accounting
    for dust creation and destruction.  ``No Dust'' has all dust
    opacities set to zero.  ``Max Dust'' assumes that all gas in the
    simulation has a Milky Way dust-to-gas ratio at all times.  The
    high mechanical feedback efficiency in this simulation dominates
    over all other concerns and the presence or absence of dust does
    not make a significant difference.}
\label{fig:mbh-a0}
\end{figure}

\begin{figure}
  \includegraphics[width=0.65\textwidth]{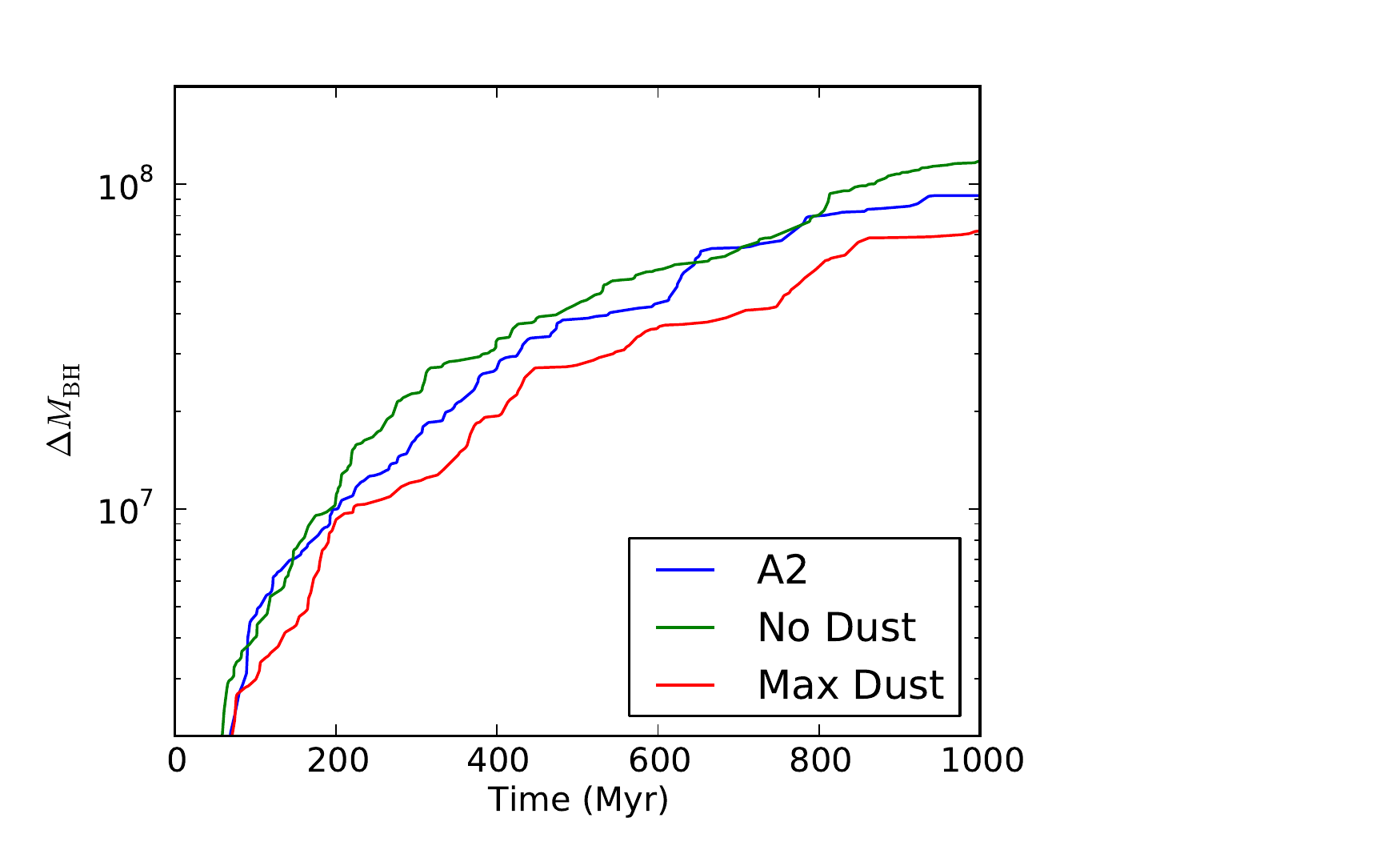}
 \caption{Black hole growth over 1 Gyr for the A2 simulation.  The
   presence or absence of dust makes up to a factor of two difference
   in the black hole growth.  }
\label{fig:mbh-a2}
\end{figure}

Figures \ref{fig:mbh-a0} and \ref{fig:mbh-a2} show black hole growth
over 1 Gyr for the A0 and A2 simulations with mechanical feedback
efficiencies of $5\times10^{-3}$ and $10^{-4}$ respectively.  In the
former case, dust has little effect on the dynamics of the simulation
because the mechanical feedback dominates over everything else.  In
the latter case, dust makes some difference, although it is not a
dramatic effect.  Furthermore, Figure \ref{fig:mbh-a2} shows that
occasionally the simulations with different assumptions about dust
arrive at very similar black hole masses in spite of the black hole
masses having been different at earlier times (e.g. at 200 Myr).  This
indicates that sometimes the dust affects {\em when} gas reaches the
black hole, but not {\em if} gas reaches the black hole.

In our simulations, dust is not negligible in determining black hole
growth, but neither is it dominant.  Extreme assumptions about the
dust-to-gas ratio ranging from no dust at all to constant Milky Way
ratios makes at best a factor of two difference in the mass added to
the black hole.

\begin{figure}
  \includegraphics[width=\columnwidth]{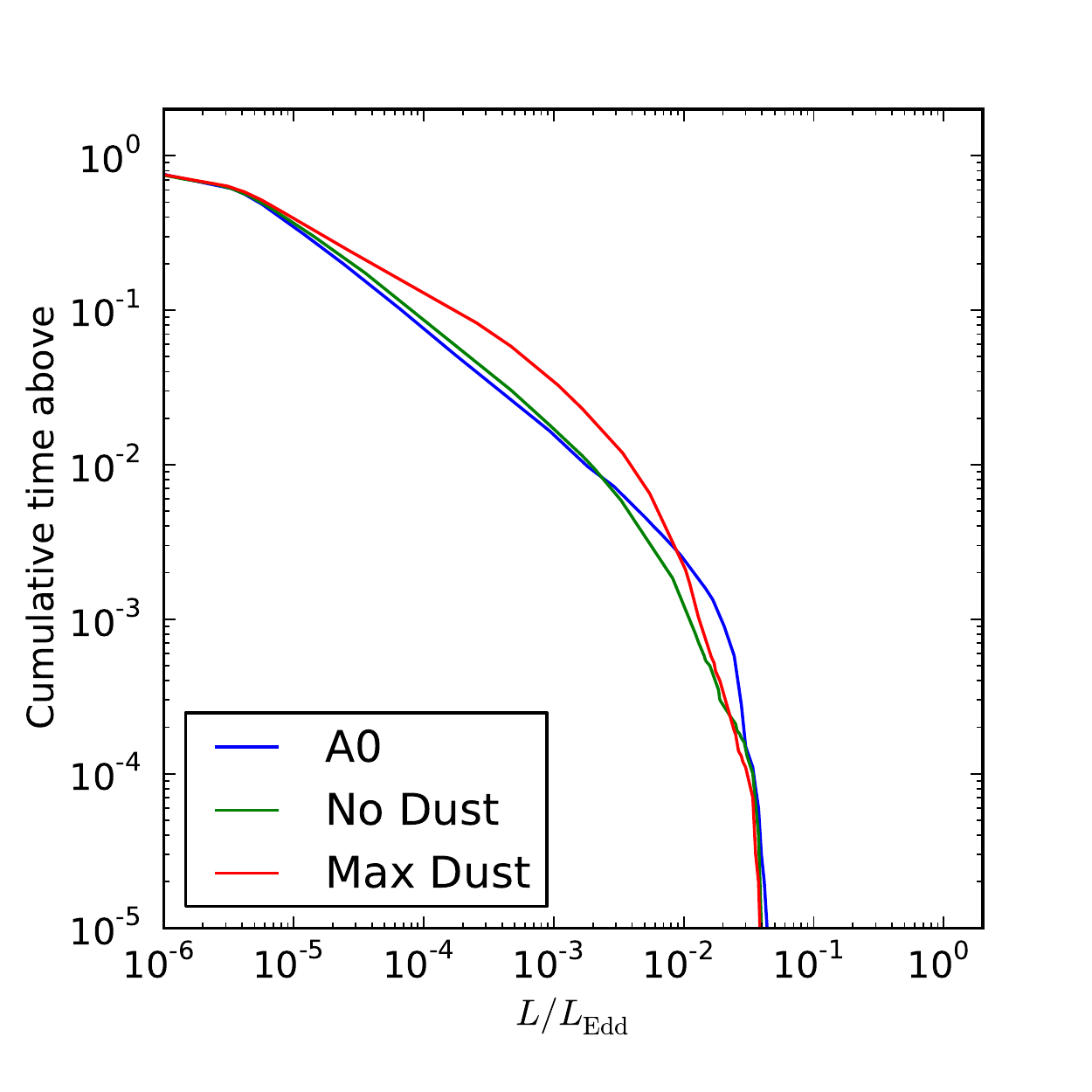}
  \caption{Cumulative time (in units of total elapsed time) where the
    ratio of the black hole accretion rate to the Eddington rate is
    above the given value for the A0 simulation with different
    assumptions about the dust to gas ratio.  Dust does not make a big
    difference, although the ``Max Dust'' simulation spends somewhat
    more time with an Eddington ratio of $10^{-4}$ to $10^{-2}$
    compared to the other simulations.}
 \label{fig:edd-a0}
\end{figure}

\begin{figure}
  \includegraphics[width=\columnwidth]{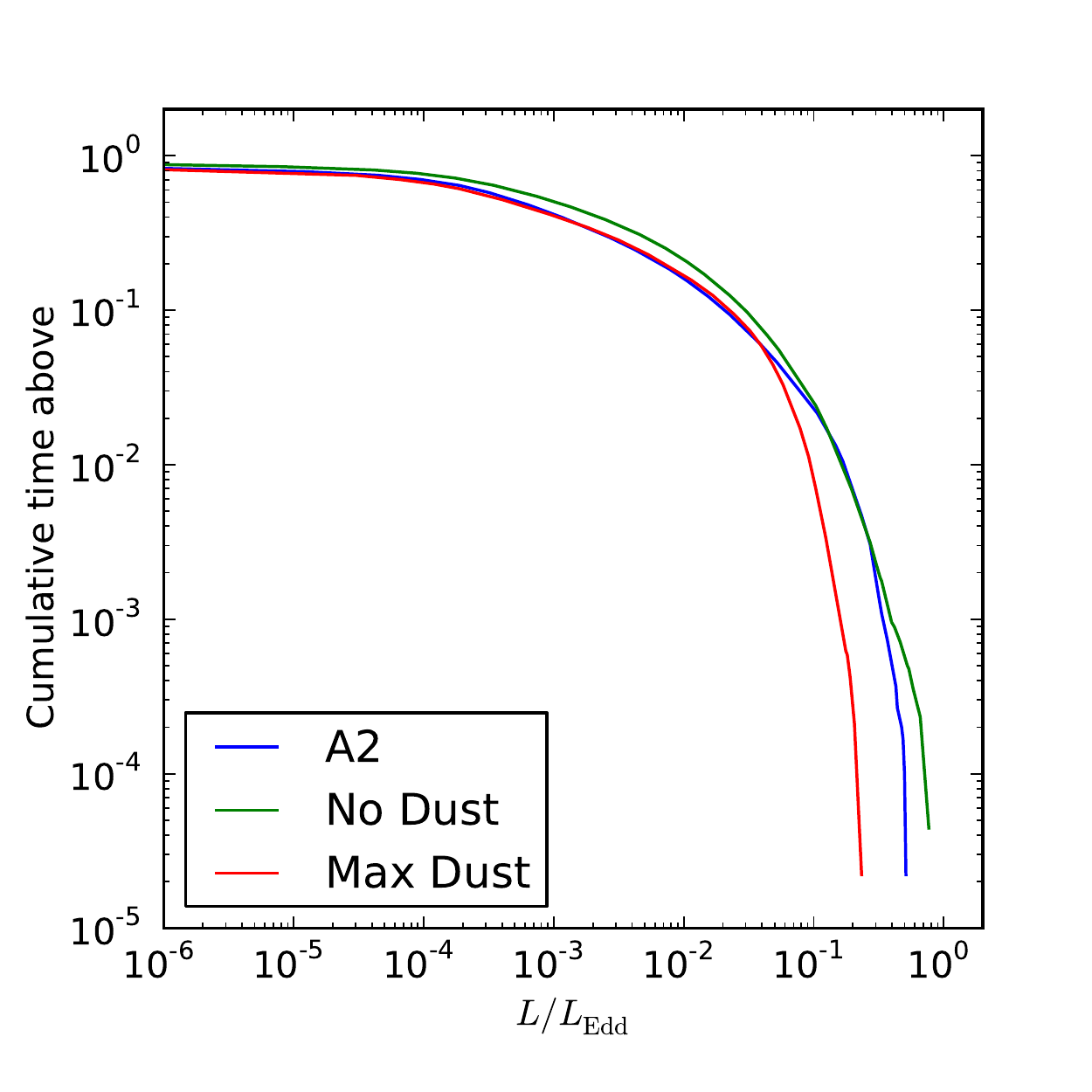}
 \caption{Cumulative time where the ratio of the black hole accretion
   rate to the Eddington rate is above the given value for the A2
   simulation with different assumptions about the dust to gas
   ratio. Radiative forces on dust grains apparently prevent the black
   hole from reaching Eddington ratios greater than about 20\%.  }
 \label{fig:edd-a2}
\end{figure}

Figures \ref{fig:edd-a0} and \ref{fig:edd-a2} show the cumulative
distribution of the ratio of the black hole accretion rate to the
Eddington rate.  The differences are not huge, but in the A2 case,
dust opacity apparently does prevent the black hole from reaching the
Eddington rate: the maximum is about 20\% of the Eddington rate.  The
fraction of the time when the SMBH would be identified as a quasar
($L/L_{\rm Edd} > 0.1$) is moderate.

\begin{figure*}
  \includegraphics[width=1.0\textwidth]{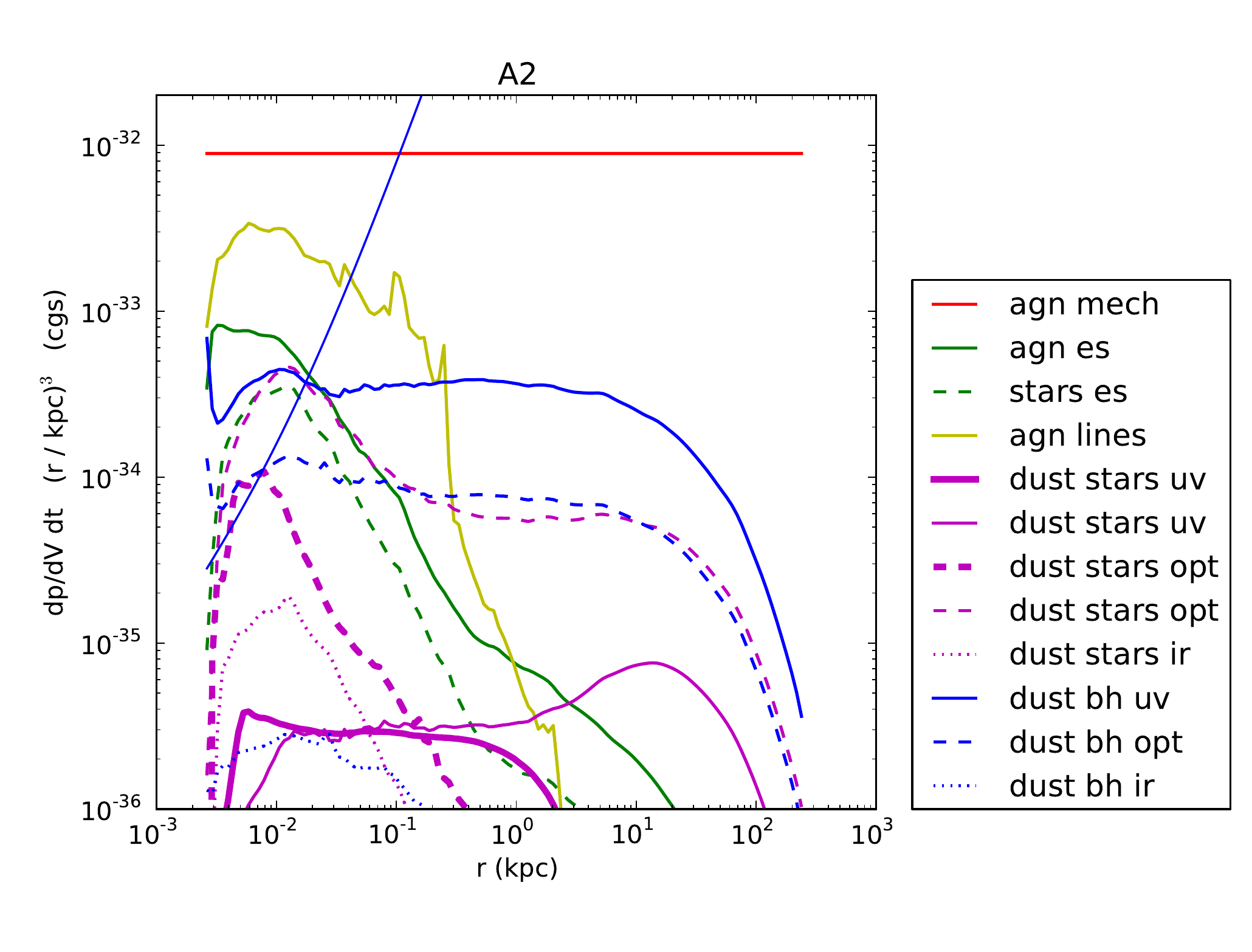}
  \caption{Time averaged radial momentum source and sink terms over 1
    Gyr for the A2 simulation.  Each line shows the time averaged
    force per unit volume imparted by each physical process as a
    function of radius.  Curves have been multiplied by
    $(r/\mbox{kpc})^3$ to flatten them out.  The thin rising blue line
    is gravity.  Heavy lines are forces directed inward.  Where a
    given physical process has both a heavy and light line, the force
    is sometimes directed inward and sometimes directed outward.  The
    red line is for mechanical momentum.  Mechanical momentum is
    injected at the first grid cell, so the quantity plotted is the
    force per unit volume if the momentum is deposited uniformly
    within a sphere of radius $r$.  The green lines are for electron
    scattering of photons produced by the AGN or by stars, the yellow
    line is for photoionization by the AGN, the pink lines are for
    dust opacity for photons produced by the stars in different bands,
    and the blue lines are for dust opacity for photons produced by
    the AGN in different bands.  The mechanical momentum accounts for
    the bulk of the momentum injected by feedback, and it will
    dominate at the radius at which the momentum is actually deposited
    into the ISM (typically $\simeq$1 kpc).  Photoionization by the
    AGN (AGN lines) is important inside a few hundred parsecs, outside
    of which AGN UV photons acting on dust are important.  }
  \label{fig:momentum}
\end{figure*}

\begin{figure*}
  \includegraphics[width=1.0\textwidth]{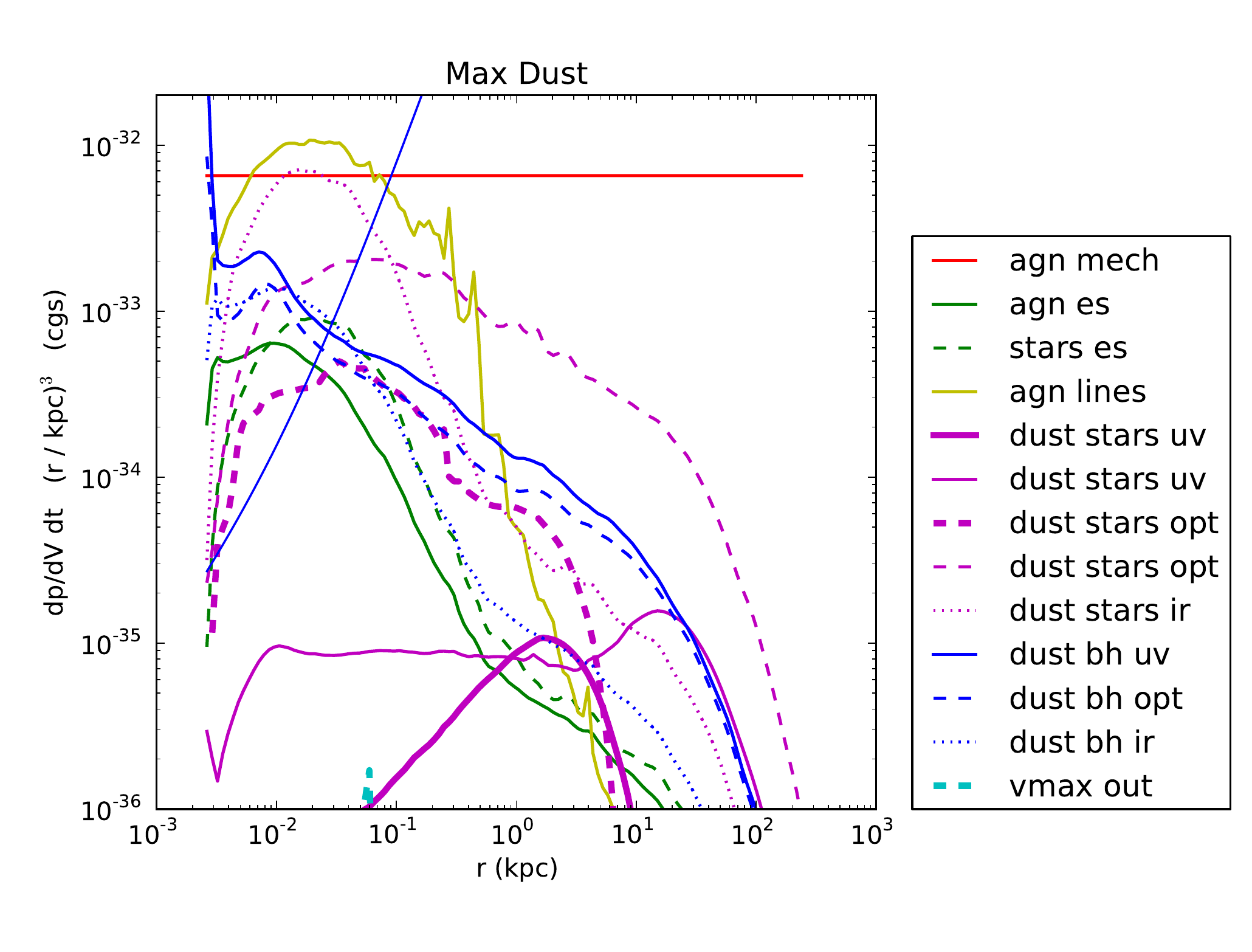}
 \caption{Radial momentum source and sink terms over 1 Gyr for the A2
    simulation, assuming a Milky Way dust-to-gas ratio at all times.
    The lines and labels are the same as Figure \ref{fig:momentum}
    except for the cyan line, which shows the very small amount of
    momentum removed by limiting the gas velocity to $30,000 \unit{km}\unit{s}^{-1}$. 
    The most important momentum sources are the mechanical wind,
    photoionization by the AGN, and scattering of stellar IR photons
    by dust grains within about 1 kpc.  Outside of 1 kpc, absorption
    of optical stellar photons by dust grains is the dominant source
    of momentum.}
  \label{fig:momentum-max-dust}
\end{figure*}

Figures \ref{fig:momentum} and \ref{fig:momentum-max-dust} show the
effect of various physical processes on the momentum balance of the
ISM.  These figures immediately show which physical processes dominate
and which are negligible.  It is important to note that these figures
show the time averaged force per unit volume, so the dominant forces 
could change in the midst intense bursts of AGN accretion and star
formation.  

Most of the forces act on each grid cell separately, making it easy to
tabulate how much momentum each separate process imparts to the ISM.
Mechanical momentum is different: it is injected at the first grid
cell and then the conservation equations of hydrodynamics transport
that momentum outward until it mixes with the ISM and the wind stops.
However, the wind termination is not abrupt and there are significant
radial motions in the ISM even in the absence of a mechanical wind
from the AGN, so determining the exact radius at which the mechanical
wind deposits momentum is somewhat ambiguous.  The quantity plotted as
the red line in Figures \ref{fig:momentum} and
\ref{fig:momentum-max-dust} is the force per unit volume if the
momentum is deposited uniformly within a sphere of radius $r$.  Since
the lines are also multiplied by $(r/\mbox{kpc})^3$, mechanical
momentum shows up as a straight horizontal line.  At very small radii,
the ``dentist drill'' effect ensures that not much radial momentum is
actually mixed into the ISM.  The mechanical wind typically does not
extend to the outer edge of the simulation grid, so likewise little
momentum is actually deposited at large radius.  Therefore the curve
showing where the mechanical momentum is {\em actually} deposited
would be low at small radii, low at large radii, nearly equal to the
red line at intermediate radii (typically about 1 kpc) where the
radial momentum is from the wind is actually mixed into the ISM.

For the A2 simulation with dust destruction allowed, the most
important sources of radial momentum are the mechanical wind,
photoionization by AGN photons, and absorption of UV AGN photons by
dust grains.  When dust destruction is not allowed and therefore the
effects of dust are maximized, Figure \ref{fig:momentum-max-dust}
shows that the most important sources of momentum are the mechanical
wind, photoionization by AGN photons, and dust scattering and
absorption of optical and IR photons originating from stars.  In the
latter, maximal dust case, the effect of stellar photons on dust
grains comes becomes more important than the effect of AGN photons on
dust grains for three reasons: 1) the black hole mass is about a
factor of two smaller, so the black hole produces fewer photons; 2)
about 20\% more stars are formed, so stellar luminosity is increased; 
3) the stars that form are {\em much} more centrally
concentrated (about a factor of two more stars inside of 1 kpc), so
the outgoing stellar photons see a larger optical depth to dust
absorption because the density of the ISM falls steeply with radius

It is somewhat surprising that dust seems to have such a small effect
on the black hole growth in the simulation, given that the opacity in
the UV due to dust is 3500 times the electron scattering opacity.
Therefore, at least in the ``Max Dust'' case where a Milky Way dust to
gas ratio is assumed for all gas, one might expect that the effective
limit to the black hole accretion rate is $\sim$1/3500 times the
Eddington rate.  However, the crucial crucial difference is that
radiative transfer is governed by the {\em absorption} opacity in the
UV while electron scattering and the transfer of thermal IR photons is
a {\em scattering} problem.  Once the UV has been absorbed, the
radiative transfer in the IR is, again, essentially a scattering
problem with $\kappa_{\rm IR} = 5.71 \kappa_{\rm ES}$; however, dust
to gas ratios in elliptical galaxies are depleted as compared to the
Milky Way by more than the modest factor of five required for electron
scattering to dominate over dust scattering.  

As noted by \citet{thompson:05} and \citet{murray:05} in the context
of this problem, the force exerted on a shell of gas by photons where
scattering dominates is:
\begin{equation}
  \force = \frac{L \opticaldepth}{c} \, ,
  \label{eq:thompson-force}
\end{equation}
where $L$ and $\opticaldepth$ are the luminosity of the source and
optical depth of the shell, both defined in the given frequency band.
The energy in photons cannot be destroyed, so if the shell is
optically thick, then photons build up until the gradient of the
photon number density is sufficient for diffusion to carry the input
luminosity.  This is why the Eddington luminosity effectively limits
the black hole accretion rate even for Compton-thick sources, with the
caveat that some energy is transferred to thermal energy in the gas by
inelastic scattering of photons with energies of order 1 MeV.  The
Eddington limit remains the relevant limit until the optical depth to
electron scattering becomes so large that the photon diffusion time is
larger than the gas inflow time, in which case the photons are
advected into the black hole along with the gas in a radiatively
inefficient accretion flow.

However, the situation is quite different when the total opacity is
dominated by absorption.  For UV photons emitted by the black hole,
the force exerted on a shell of gas is:
\begin{equation}
  \force = \frac{L \min(\opticaldepth, 1)}{c} \, .
\end{equation}
The incident UV photons deliver their momentum to the shell, after
which they are converted to thermal IR photons where the opacity is
much smaller.  The thermal IR photons then simply leave the galaxy
without further scattering until the galaxy becomes optically thick in
the IR.  The crucial point is that UV photons cannot build up and then
diffuse out of the galaxy, as is the case when considering the
electron scattering opacity and the dust opacity in the IR.  The
maximum force that can be exerted on a shell of gas is $f_{\rm UV}
L/c$ where $f_{\rm UV}$ is the fraction of the black hole
bolometric luminosity that is emitted in the UV.

If the infalling clouds of gas are optically thin in the UV, then the
force on a gas cloud is
\begin{equation}
  \force = \frac{L (f_{\rm UV} \tau_{\rm UV} + \tau_{\rm ES})}{c} \simeq
  \frac{L \tau_{\rm ES}}{c}
  \frac{f_{\rm UV} \kappa_{\rm UV}}{\kappa_{\rm ES}} \, , 
\end{equation}
where $L$ is the bolometric luminosity and $f_{\rm UV}$ is the
fraction of the bolometric luminosity emerging in the UV.  The cloud
will feel a force much larger than if the dust opacity were not
considered.  However, if the cloud is optically thick in the UV (but
not in the IR) then the force is
\begin{equation}
  \force = \frac{L (f_{\rm UV} + \tau_{\rm ES})}{c} \, .
\end{equation}
In this case, the force is nearly independent of the mass of the
cloud; adding mass to an already optically thick cloud increases the
gravitational force but not the opposing radiative force.  This fact
makes it difficult for the dust opacity in the UV to have a dramatic
effect on the gas falling toward the black hole.

\begin{figure*}
  \centering
  \includegraphics[width=0.49\textwidth]{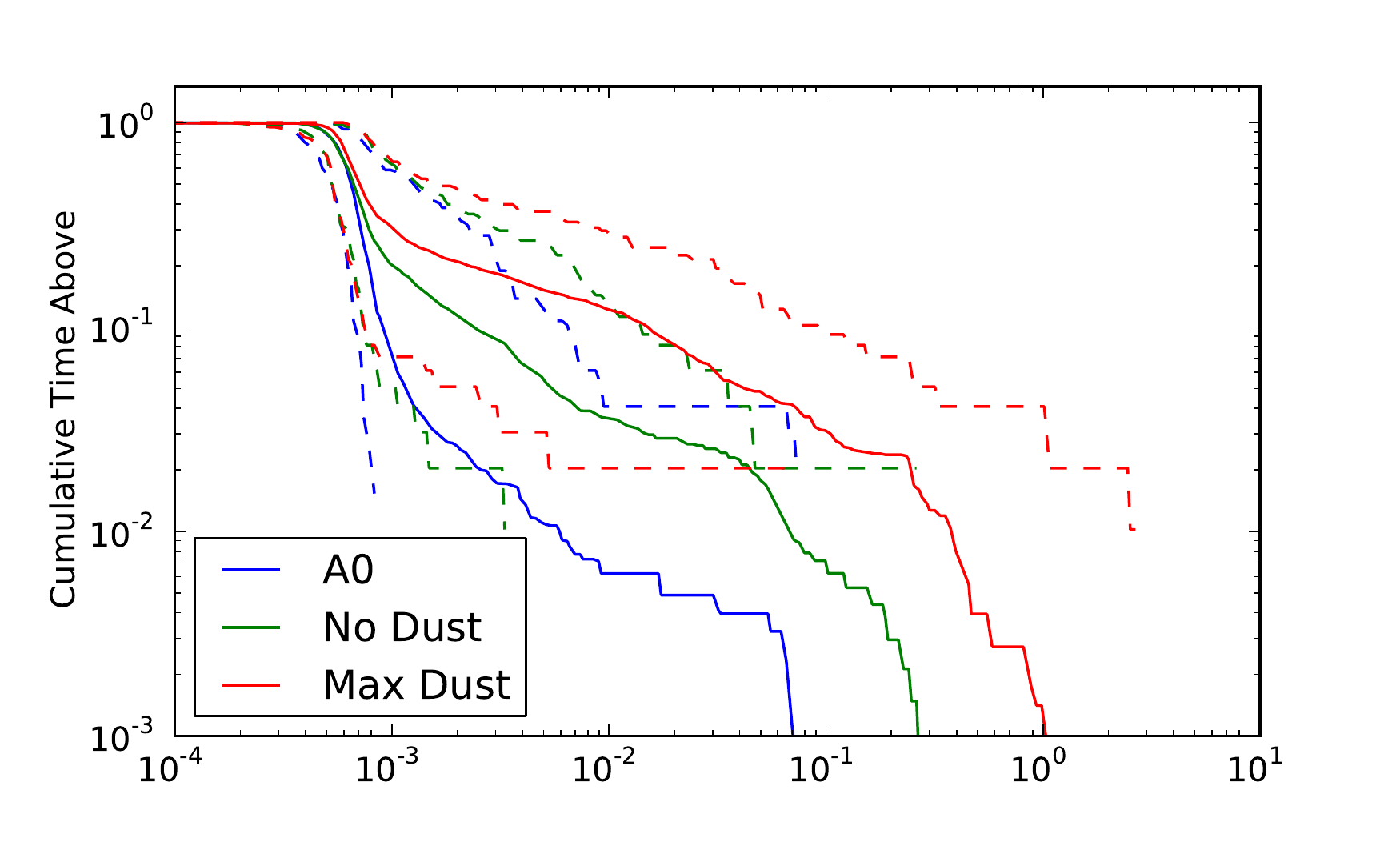}
  \includegraphics[width=0.49\textwidth]{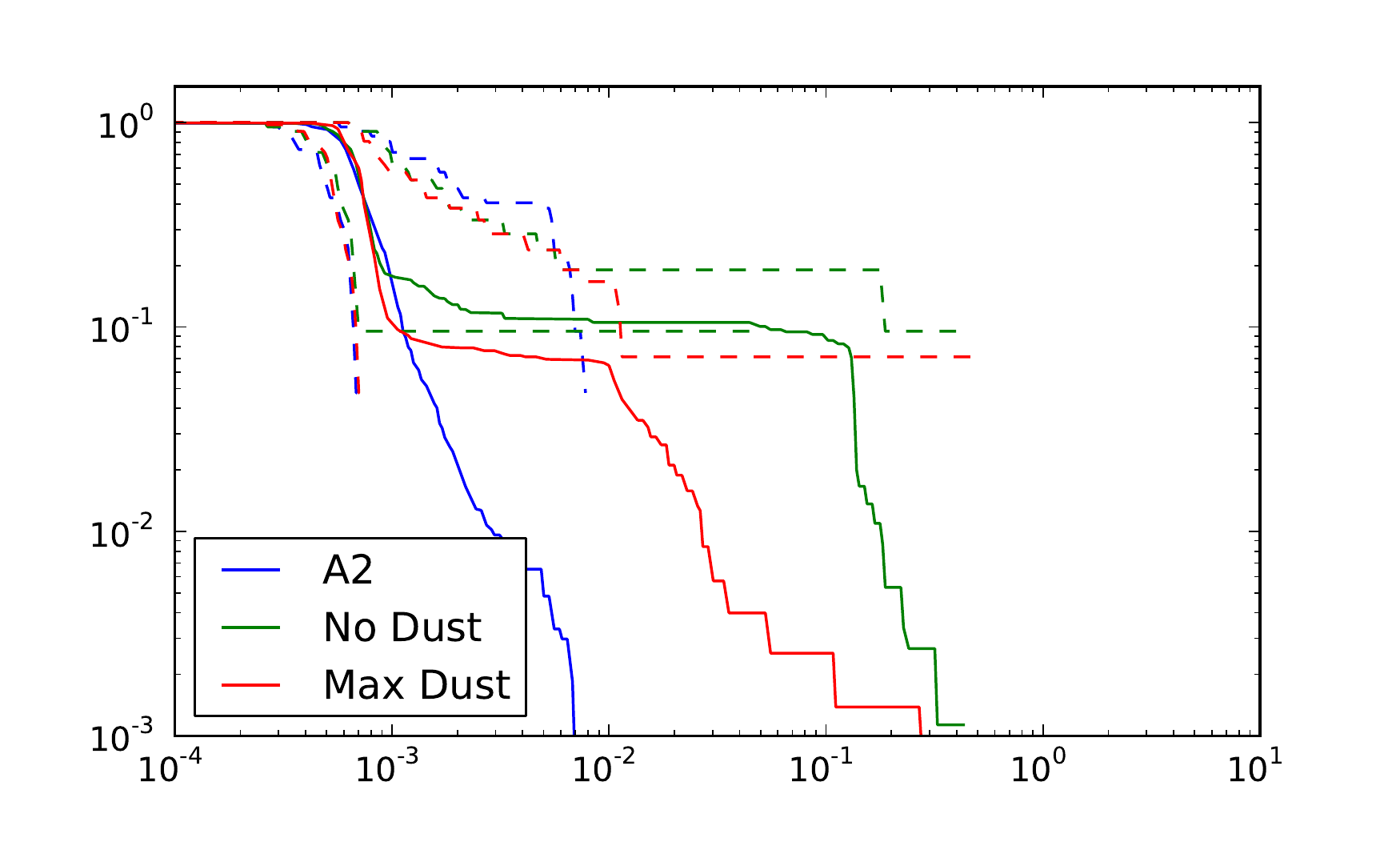}
  \includegraphics[width=0.49\textwidth]{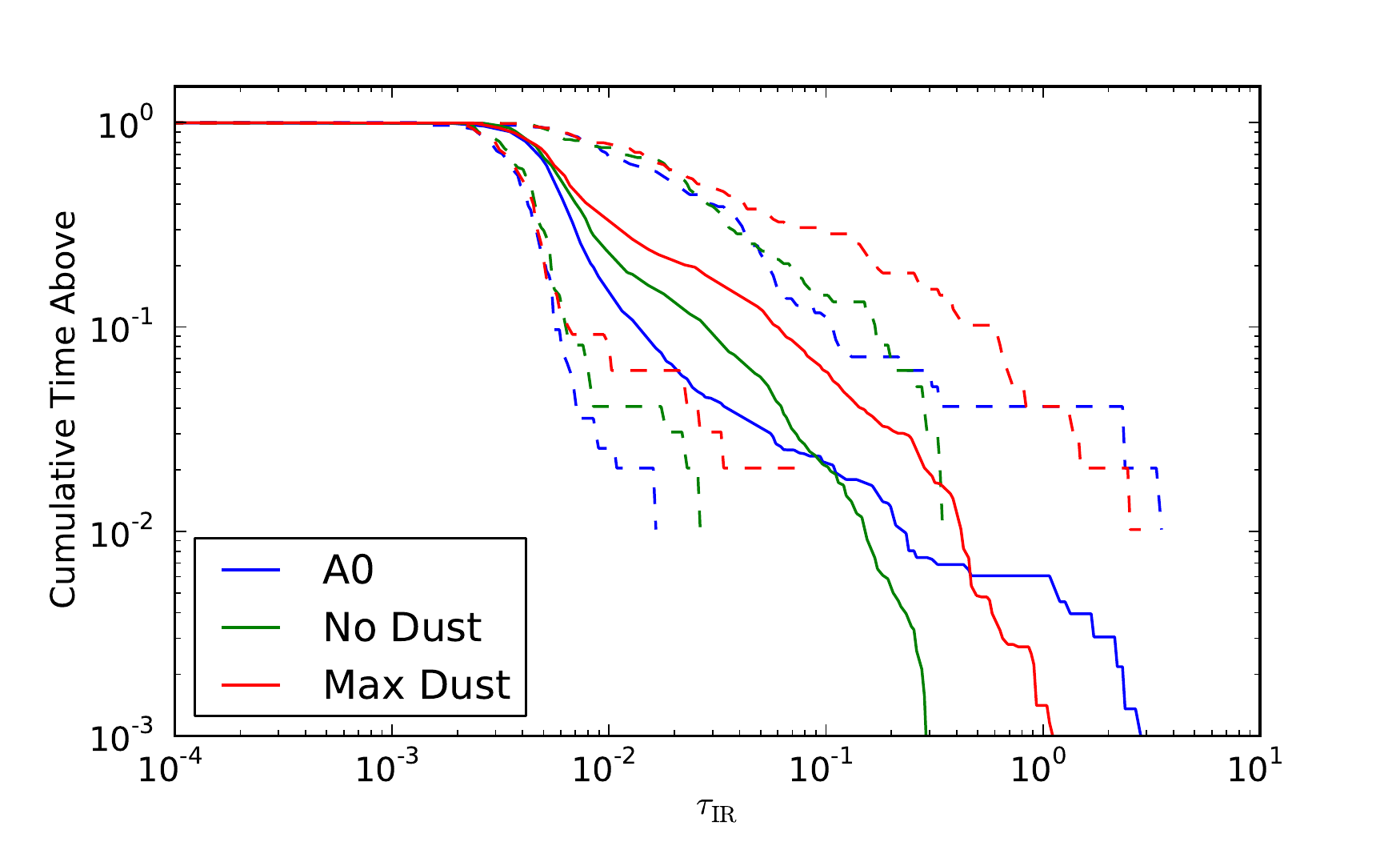}
  \includegraphics[width=0.49\textwidth]{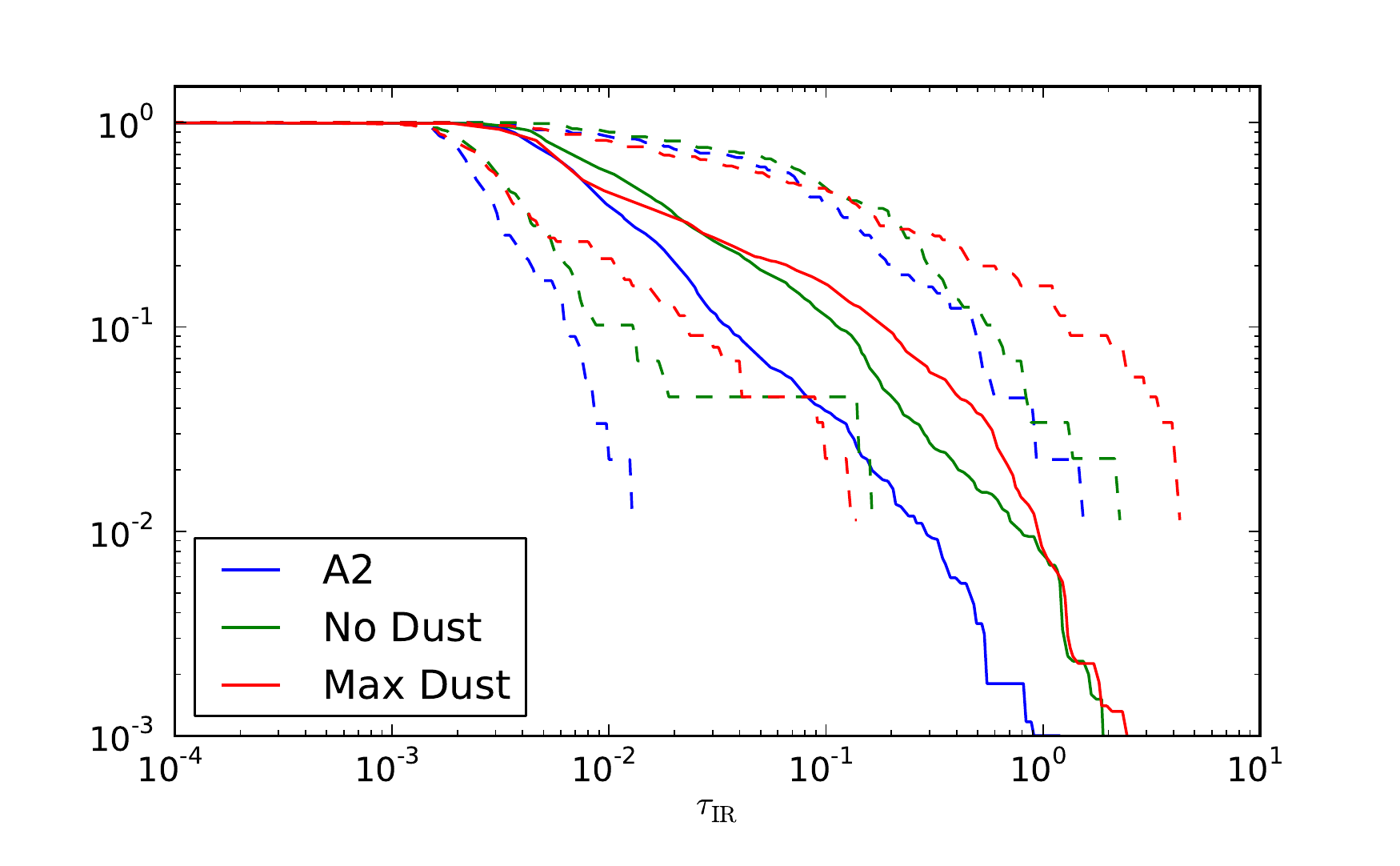}
  \caption{In the upper two panels, the cumulative distribution of
    optical depths due to dust in the IR using the tracer field that
    tracks dust creation and destruction in the simulation to
    calculate the dust opacity.  The lower two panels show the same
    quantity assuming a constant dust-to-gas ratio (corresponding to
    values appropriate for the Milky Way) so that the optical depth
    scales directly with the column density of the gas.  For solid
    lines, each line of sight in each snapshot contributes separately
    to the distribution.  Dashed lines show the distributions where
    only the line of sight with the highest or lowest optical depth
    from each snapshot contribute.  The simulations only rarely become
    optically thick in the IR, and when this happens, the optical
    depth is only a few times unity.  Furthermore, the distribution of
    minimum optical depths shows that there are always optically thin
    lines of sight available that allow photons to exit the galaxy
    without scattering multiple times.  For the simulations labeled
    ``No Dust,'' the simulation neglects any radiative forces
    associated with dust, but it nevertheless tracks the dust tracer
    field, allowing those simulations to be included in these plots.}
  \label{fig:tau-dist}
\end{figure*}

Figure \ref{fig:tau-dist} shows the distribution of optical depths for
the A0 and A2 simulations and variants.  The typical optical depths
are far below unity, although a small fraction of the time the
simulations become optically thick, at least along some lines of
sight.  However, the distribution of minimum optical depths for each
simulation snapshot show that there are almost always optically thin
lines of sight available that will allow IR photons to escape without
experiencing multiple scatterings that are necessary for
\eqnref{eq:thompson-force} to be relevant.

In spite of dust having at best a minor effect on the dynamics of the
ISM, it has a dramatic effect on the radiation emerging from the
galaxy.  With no dust grain destruction, the optical depth in the UV
from the center of the galaxy is typically of order unity.  This means
that the typical UV photon from the black hole is absorbed on its way out
of the galaxy.  Therefore {\em observations} of the AGN will be
dramatically different for different assumptions about the creation
and destruction of dust, even if the dust does not dramatically affect
the dynamics of the galaxy.

Absorbed UV/optical photons from the AGN are converted to IR photons,
but the galaxy is typically quite optically thin in the IR, so these
photons escape without much fanfare.  If the ISM {\em were} optically
thick to IR photons, then they could have a dramatic effect as they
scatter out of the galaxy, as pointed out by \citet{murray:05} and
\citet{thompson:05}, but we do not find the conditions necessary for
this effect to prevail in our simulations.  We find that in our
detailed treatment of the radiative transfer, the factor
$\opticaldepth$ in \eqnref{eq:thompson-force} never exceeds unity by a
significant amount.  

Most stellar UV/optical photons are {\em not} absorbed because the
star light is emitted at larger radius and need not traverse the
densest part of the ISM.  The typical stellar UV photon is produced at
roughly the half-light radius of the galaxy, from which point the
typical optical depth is far less than unity.  Therefore the stellar
emission is largely unaffected by dust in the galaxy.  Dramatically
different assumptions about the dust to gas ratio have a minor effect
on both the dynamics due to and the observations of stellar photons.

A subtlety arises when considering the distributed radiation source
(the stars) in the case where the ISM is optically thick to absorption
and the gas density is falling steeply (faster than $r^{-1}$).  In this
case it is possible for the luminosity at a given radius to be
negative, i.e. the net flux points inwards.  If there were no
absorption, then every ingoing photon would eventually become an
outgoing photon, either by traveling unimpeded and emerging at a
different point on the same sphere of radius $r$, or else by
scattering until it diffused to radius $r$.  This must be the case or
else photons would build up at radii smaller than $r$.

However, when absorption is important, UV photons may propagate
inward, undergo absorption, and re-emerge as IR photons.  If
the density profile is steep enough, this situation robustly leads
to negative UV luminosities for radii smaller than the radius from
which the typical UV photon is emitted.  That is to say that the stars
at larger radius exert a force acting to {\em compress} gas near the
center of the galaxy.  The force is {\em not} balanced by outgoing IR
photons (which must after all carry the energy of UV photons absorbed
at small radius) because the opacity is smaller in the IR.  The
momentum carried by the IR photons is deposited at larger radius.

We find that this is often the case in our simulations: the gas
density profile is generally steep enough to permit negative
luminosities to develop within a few kiloparsecs of the center of the
galaxy.  This effect can be seen in Figures \ref{fig:momentum} and
\ref{fig:momentum-max-dust}.

There are situations in which dust plays a major role, such as
galaxies undergoing starbursts or merger-induced quasar activity with
copious quantities of cold gas \citep[][]{murray:05, thompson:05,
  debuhr:10, debuhr:11}.  We have performed one simulation designed to
be similar to an ultra-luminous infra-red galaxy having a star
formation rate of several hundred solar masses per year.  The
simulation uses the A2 feedback parameters but starts with
$3\times10^{10} M_\odot$ of gas, corresponding to a gas fraction of
10\%, as could be induced by a major merger 
or an inflow of cold gas. 
The star formation rate reaches peaks of 200 solar masses per year,
approaching the ULIRG conditions.

\begin{figure}
  \includegraphics[width=\columnwidth]{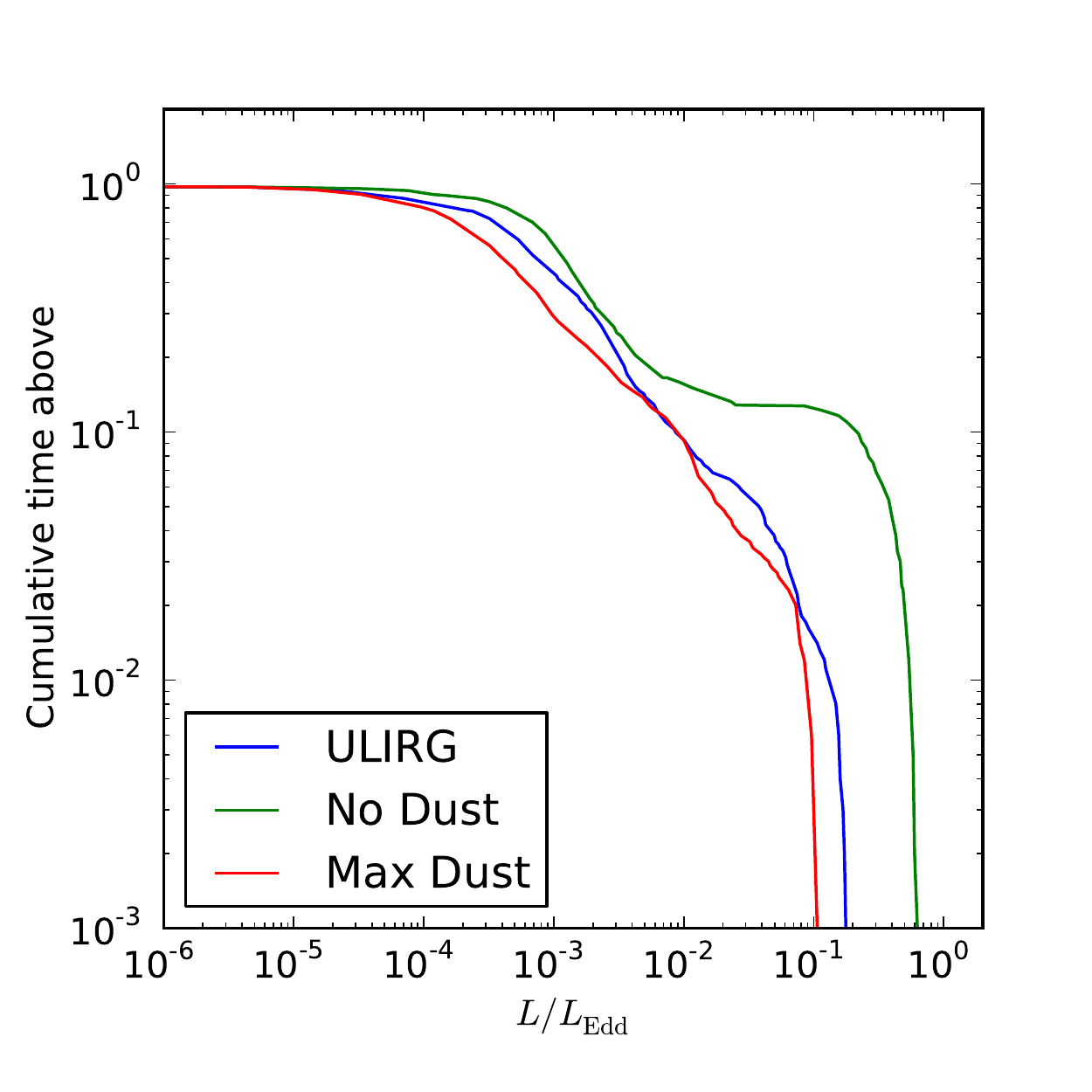}
  \caption{Cumulative time where the ratio of the black hole accretion
    rate to the Eddington rate is above the given value for a
    simulation designed to resemble a ULIRG with different assumptions
    about the dust to gas ratio. For large enough gas column
    densities, radiative effects due to dust opacity have a large
    effect, and a careful treatment of dust production and destruction
  produces results similar to the ``Max Dust'' case, with ``ULIRG'' SMBH
  luminosity in the quasar range for several percent of the time.}
 \label{fig:edd-ulirg}
\end{figure}

Figure \ref{fig:edd-ulirg} shows the distribution of Eddington ratios
for different assumptions about the dust to gas ratio in this
simulation.  The simulation that includes dust creation and
destruction is quite similar to the maximum dust case because the
short dust creation time ensures that most of the gas has a dust to
gas ratio near that of the Milky Way, as is observed in the ULIRG
case.  Assuming that there is {\em no} dust in the simulation yields a
dramatically different distribution of Eddington ratios, reaching much
larger values.  
The time averaged black hole accretion rates are 0.045
$M_\odot \unit{yr}^{-1}$ for the case with maximal dust, 0.061 $M_\odot \unit{yr}^{-1}$ for
the case with dust creation and destruction, and 0.32 $M_\odot \unit{yr}^{-1}$
for the case with no dust.  Thus the presence of dust may suppress
black hole growth by a factor of five to seven depending on the
assumptions surrounding dust creation and destruction.
 
The reason for this is that the conditions now resemble those
envisioned by \citet{thompson:05} and \citet{murray:05}: the gas
surface density is high enough that the galaxy is optically thick in
the infrared.  The infrared photons build up in the galaxy until the
gradient of the photon number density is sufficient to carry the black
hole luminosity by photon diffusion.  In this case, the relevant
opacity to limit accretion is not the electron scattering opacity but
the dust opacity in the infrared.  Dust then has a dramatic effect on
the gas dynamics and on the growth of the black hole.

\begin{figure}
  \includegraphics[width=\columnwidth]{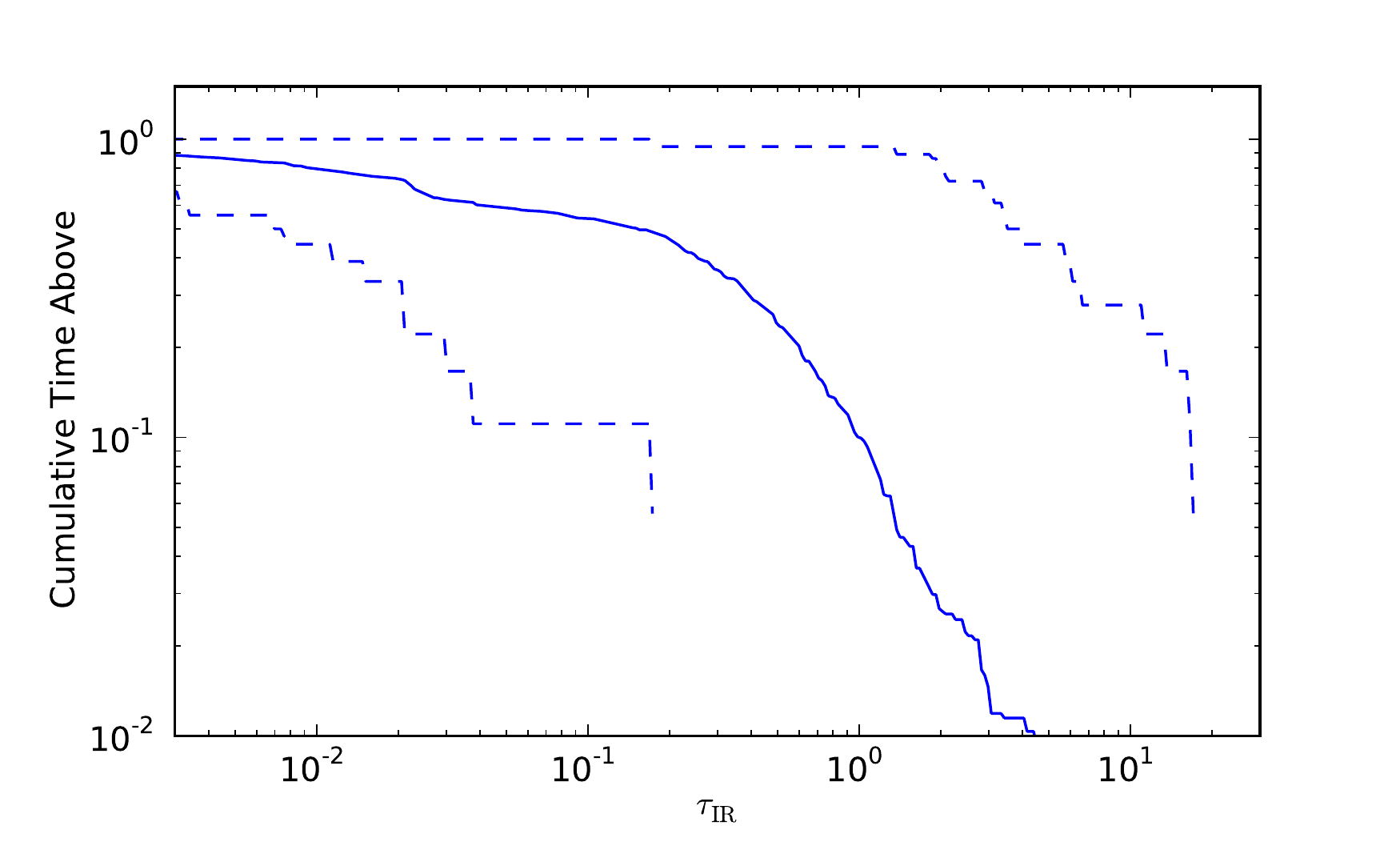}
  \caption{Cumulative distribution of optical depths due to dust in
    the IR for the simulation designed to resemble a ULIRG.  The ``Max
    Dust'' case is shown, but high gas densities resulted in
    sufficiently fast dust creation that the other cases do not differ
    significantly.  Typical optical depths along a typical line of
    sight are of order unity.  There are almost always optically thick
  lines of sight where the optical depth is larger than ten a significant
  fraction of the time.  However, even in this case there are always
  optically thin lines of sight that would allow photons to easily
  escape from the galaxy.}
 \label{fig:tau-dist-ulirg}
\end{figure}

Figure \ref{fig:tau-dist-ulirg} shows the cumulative distribution of
optical depths due to dust in the IR for the ULIRG-like simulation.
Along some lines of sight, the situation resembles that envisioned by
\citet{thompson:05} and \citet{murray:05} a significant but
sub-dominant fraction of the time---optical depths are a few times
unity and can exceed ten.  However, the distribution of minimum optical
depths for each simulation snapshot shows that IR photons would always
be able to escape via optically thin ``windows''.  In order for
equation \ref{eq:thompson-force} to pertain, the IR photons must be
effectively trapped in the galaxy so that the only way for them to exit
the galaxy is via diffusion.  If there are optically thin lines of
sight, this will not occur.

\section[]{Discussion and Conclusions}
\label{sec:discussion} 

We have outlined a method for computing the forces on interstellar gas
due to absorption of photons by dust grains.  The radiative transfer
equation is recast into a set of differential equations by taking
moments in the photon propagation direction.  The set of differential
equations comprises a boundary value problem, the solution of which
gives the photon field which can be used to compute forces on the
ISM.  We identified a numerical method that reliably gives the
solution to the set of differential equations even for very large
optical depths.  

The method gives the correct asymptotic behavior in all of the
relevant limits (dominated by the central point source; dominated by
the distributed isotropic source; optically thin; optically thick to
UV/optical; optically thick to IR; photon field nearly isotropic;
photon field highly directed) and reasonably interpolates between the
limits when necessary.  The method is explicitly energy conserving so
that UV/optical photons that are absorbed are not lost, but are rather
redistributed to the IR where they may scatter out of the galaxy.  It
does not require spherical symmetry, but it only handles the radial
component of radiative forces.

We have also described a physically motivated simplification of the
above method based on splitting the photon field into ingoing and
outgoing streams.  This method is much simpler to implement and shares
many of the desirable characteristics of the more complicated method.  

We have implemented both of these numerical methods in a hydrodynamic
code with source terms for energy, momentum, and mass appropriate to
study black hole fueling and star formation in early type galaxies.
Both radiative and mechanical feedback are included for both the AGN
and the stars.  We allow for dust destruction by sputtering and dust
creation by supernovae, planetary winds, and optionally gas-phase
grain growth.  

In the case of secular fueling of a normal SMBH in an elliptical galaxy,
the dynamics of the simulations are not greatly affected by the
presence or absence of dust.  At high AGN mechanical feedback
efficiencies, dust has very little effect because mechanical feedback
dominates.  At lower mechanical feedback efficiencies, dust has some
effect but large changes in our assumption about the dust-to-gas ratio
lead to measurable but relatively modest changes in the black hole
growth, star formation, and galactic winds.  This is true even for
extreme assumptions about dust, ranging from no dust at all to
assuming dust-to-gas rations appropriate to the Milky Way pertain at
all times.  Allowing for the possibility of dust destruction by
sputtering in hot gas makes the simulations behave similarly to the
``No Dust'' case.

If the galaxy is optically thick to IR photons, repeated scattering of
these photons can have a dramatic effect on the momentum balance of
the ISM.  We have verified this by simulating a galaxy with a very
large mass in the ISM, as might follow a major merger with a gas rich
galaxy, in which case the presence or absence of dust
makes a large difference, as expected.

However, we do not find that this situation is typically realized in
our simulations.  The opacity of dust in the IR is about five times
the electron scattering opacity, so a system must be Compton thick
before this scattering starts to have a large effect.  Our simulated
galaxies are occasionally Compton thick, but no more, and they
typically have much lower column densities.  Even with no dust
destruction, we find that the optical depth in the IR is at most a few
times unity, and it reaches that value only for a small fraction of
the total simulation time.  We find that it is very rare for the optical depth
in the IR to significantly exceed unity, as is assumed by \citet{debuhr:10,
  debuhr:11}.  When dust destruction is allowed, the optical depths in
the IR is typically smaller by several orders of magnitude.

\section*{Acknowledgments}

We thank Bruce Draine, Daniel Proga, and Sergey Sazonov for useful
discussions.  G.S.N. was supported by the Princeton University Council
on Science and Technology and made extensive use of the computing
facilities of the Princeton Institute for Computational Science and
Engineering.  L.C. is supported by the MIUR grant CoFin08.

\bsp

\appendix

\section[]{Third order radiation transport}
\label{sec:complex-rxfer}

The quality of a solution to the radiation transport equation obtained
by taking moments depends on the accuracy of the physical assumptions
implicit in the chosen analytic form for the specific intensity.  If
those assumptions are not satisfied, taking the expansion to higher
order is unlikely to provide benefits commensurate with the costs in
terms of complexity and computational resources.  We believe that the
assumptions embodied by \eqntworef{eq:intensity}{eq:thin-thick} are
quite reasonable for the problem at hand.  Nevertheless, it is
possible to avoid using \eqnref{eq:thin-thick} by taking one
additional moment of the specific intensity and radiation transport
equation.  We provide the equations here for reference but do not
pursue the method further in the present paper.

Define the third moment of the specific intensity to be:
\begin{equation}
  \ithree \equiv \int_0^{2\pi} d\phiprop \int_{-1}^1  \mu^3
  I(\mu,\phiprop) \, d\mu \, .
\end{equation}
Given \eqnref{eq:intensity} we find
\begin{equation}
\ithree = \frac{4\pi B}{5} + \pi D  \, .
\end{equation}
We have three constants in the assumed form of the mean intensity
and four moments of the mean intensity, so we can solve for the
closure relation:
\begin{equation}
  \ithree = \frac{4\pi J + 3 \flux + 3 c P}{5} \, .
\end{equation}

To obtain the differential equations, we take the equations for the zeroth
and first moment derived above, supplement them with the second
moment of the radiation transport equation, and then insert the
closure relation to obtain the following set of three equations:
\begin{eqnarray}
\frac{d J}{d r} 
+ \frac{7 J}{r} 
- \frac{\flux}{\pi r}
+ \frac{3 c P}{4\pi r}
= \rho \left(3 \kappa_a  + \frac{5 \kappa_s}{3} \right) J 
- \frac{\dot{E}}{3 \pi} 
& & \nonumber \\
+ \frac{\rho (\kappa_a + \kappa_s) (3 \flux - 5 c P)}{4 \pi} \, ,
\end{eqnarray}
\begin{equation}
\frac{d L}{d r} = 4\pi r^2 (\dot{E} - 4 \pi \rho
\kappa_a J) \, ,
\end{equation}
and
\begin{equation}
  c\frac{d P}{d r}
- \frac{4 \pi J}{r}
+ \frac{3 c P}{r}
= -\rho (\kappa_a + \kappa_s) \flux \, .
\end{equation}

Adopting the use of these equations and closure relation is perhaps
formally more appealing than the method outlined in Section
\ref{sec:radiative-transfer}.  
However, the cost is one additional equation to integrate
and perhaps less obvious physical content to the closure relation and the
equations themselves.  We have included the present equations for
reference, but do not use this method of obtaining solutions to the
radiation transport problem in the present work.

\section[]{Simplified radiation transport}
\label{sec:simplified-rxfer}

In Section \ref{sec:radiative-transfer} we have described a complete algorithm to solve for the
photon field and, subsequently, the radiative forces on interstellar
gas.  The algorithm is exact in the sense that if the
physical approximations used in its derivation are correct, the method
will compute the correct photon field and radiative forces.  

However, solving for the radiation field dominates the computational
resources required for the calculation primarily because the system
differential describes a boundary value problem rather than an initial
value problem and hence finding the solution requires iteration.  We
would like a {\em another} solution method that is {\em
  computationally} efficient while respecting the essential physics of
the previously defined algorithm.

Consider the problem where the only source of radiation is the black
hole and the scattering opacity is zero.  The UV/optical photons are always
radially outgoing at all times.  Some UV/optical photons are converted
to IR photons which diffuse out of the galaxy, but the UV/optical
radiation field never becomes isotropic.  In this case the flux is
simply related to the mean intensity ($\flux = 4 \pi J$) and we can
eliminate \eqnref{eq:inten} to obtain the single equation for the UV
and optical components
\begin{equation}
  \frac{dL}{dr} = - \rho \kappa_a L \, .
\end{equation}
The boundary condition is $L(r_{\rm min}) = L_{\rm BH}$.  
This is now an initial value problem rather than a boundary value
problem, and can be trivially solved by integrating from the inner
edge of the simulation grid.

This change greatly simplifies the solution procedure.  It would be
very useful to obtain a simple approximate treatment of this problem
that as an initial value problem rather than a boundary value problem.
We seek such an algorithm presently.

\subsection{Differential equations}
Conceptually, there are two streams of radiation: an ingoing stream
and an outgoing stream.  If the optical depth is low, then the photons
in the ingoing stream are likely to successfully traverse the inner
parts of the galaxy and emerge as outgoing photons.  In this case,
{\em all} of the radiation emitted by stars should be considered to be
added to the outgoing stream.  Doing so gives the correct forces on
the material inside and outside the emitting sphere due to the
well-known result that for a inverse square force law, there is no net
force on a test particle inside a spherical shell of material; and a
test particle outside the spherical shell feels a force equal to that
of a point source of equivalent mass, charge, or luminosity, no matter
whether the force is gravitational, electrostatic, or due to momentum
transfer by photons.

If the optical depth is large, then the ingoing photons are likely to
be absorbed in the inner part of the galaxy.  Then only half of the
emitted photons should be added to the outgoing stream.  The other
half should be added to the ingoing stream, where they will in due
course be absorbed.

This leads to the following two equations:
\begin{equation}
  \frac{dL_{\rm out}}{dr} = 4 \pi r^2 \Psi \dot{E} - \rho \kappa_a
  L_{\rm out} \, ,
\label{eq:lout}
\end{equation}
and
\begin{equation}
  \frac{dL_{\rm in}}{dr} = 4 \pi r^2 (1-\Psi) \dot{E} - \rho \kappa_a
  L_{\rm in} \, ,
\label{eq:lin}
\end{equation}
where $\Psi$ is the fraction of photons emitted at a given radius that
are likely to become outgoing photons, either because they were
emitted in the outward direction, or because they were emitted inward
but are likely to successfully traverse that region without being
absorbed.  

\subsection{Solving the differential equations}
The boundary conditions are $L_{\rm out}(r_{\rm min}) = 0$ and
$L_{\rm in}(r_{\rm max}) = 0$, so that \eqnref{eq:lout} is solved by
integrating outward and \eqnref{eq:lin} is solved by
integrating inward.

Summing \eqntworef{eq:lout}{eq:lin} simply gives \eqnref{eq:lum-rxfer}, as
expected.  The advantage of splitting the radiation into an ingoing
stream and an outgoing stream is that it transforms a boundary value
problem into an initial value problem, leading to a great reduction of
the effort required to find a numerical solution.  The quality of the
solution will depend on the accuracy of the function $\Psi$ is
estimating the contribution that the stars at a given radius make to
the ingoing and outgoing streams of radiation.  This can of course be
checked by comparing to the solution obtained by solving the full
boundary value problem.

\begin{figure*}
  \includegraphics[width=0.8\textwidth]{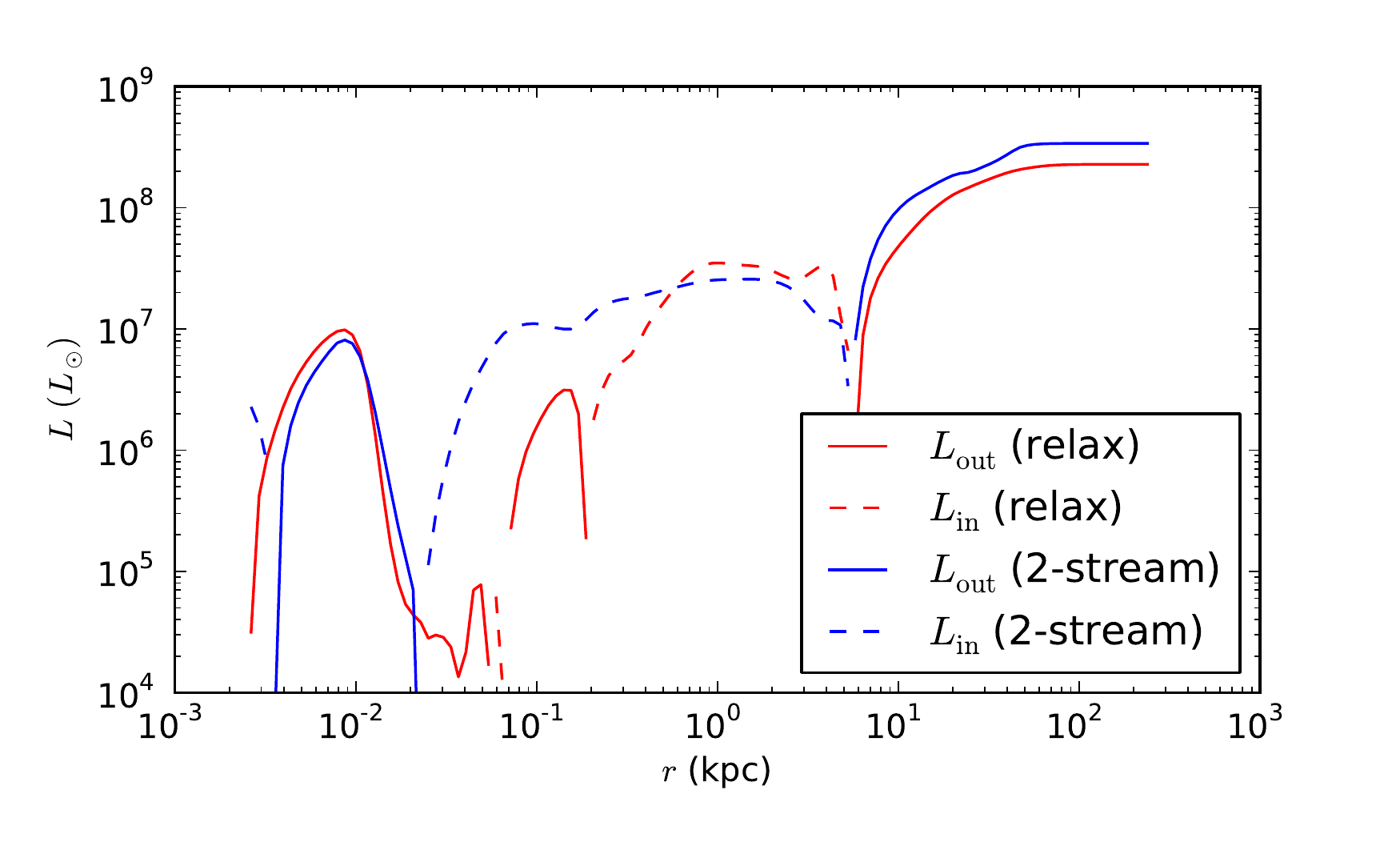}
  \caption{The luminosity as a function of radius for the full
    relaxation based algorithm (Section
    \ref{sec:mathematical-treatment}) and the simpler two-stream based 
    algorithm (Appendix \ref{sec:simplified-rxfer}).  Dashed lines
    indicate inward-directed
    luminosities.  The solution given by the two-stream approximation
    described in Appendix \ref{sec:simplified-rxfer} has about 50\%
    more flux at infinity than the ``true'' solution, and it misses
    the true solution around 100 pc.  However, the solution is quite
    good given the gains in algorithmic simplicity and reduced
    computational costs.  The function $\Psi$ associated with
    $\zeta=2$ was used for this figure, where $\zeta$ is the index of
    the assumed power law dependence of photon mean free path on
    radius used in defining $\Psi$ (see Appendix \ref{sec:psi} for
    details).}
 \label{fig:simplified-rxfer}
\end{figure*}

We have found that this scheme gives reasonable results at a fraction
of the algorithmic complexity and computational cost of obtaining the
exact solution.  Figure \ref{fig:simplified-rxfer} shows the results
of solving the radiation transfer equation by the simplified two-stream method
and the full solution obtained by solving the boundary value problem
for one line of sight in one simulation snapshot.  The total optical
depth along the line of sight is 13.5.  The solution given by the
simplified scheme is not perfect, but it is remarkably good given the
large optical depth and the reduced algorithmic and computational
cost. 

\subsection[]{Concerning $\Psi$}
\label{sec:psi}

If $\pabs$ is the probability that a given ingoing photon
is absorbed before again reaching the radius at which it was emitted
(and hence becoming an outgoing photon) and $\ptrans = 1-\pabs$ is
the probability that a given ingoing photon is transmitted so that it
again arrives at the radius at which it was emitted and becomes an
outgoing photon, then: 
\begin{equation}
\Psi \equiv  1-\frac{\pabs}{2} = \frac{1+\ptrans}{2}
\label{eq:psi}
\end{equation}
The probabilities for absorption and transmission appear in
\eqntworef{eq:lout}{eq:lin} only through the quantities $\Psi$ and
$1-\Psi$.  Note that of the four ways of writing $\Psi$ and $1-\Psi$
in terms of transmission and absorption probabilities, $\pabs$ and
$\ptrans$ are added to an order-unity constant in all cases except
one: $1-\Psi = \pabs/2$.  An argument could be made that one should
take special care to obtain the correct asymptotic behavior for
$\pabs$ when $\pabs\ll1$ in order to avoid large fractional errors in
the ingoing radiation stream.  However, this is the optically thin
limit, exactly where the ingoing radiation stream is expected to be of
limited importance.  Therefore any expression for $\Psi$ with the
correct asymptotic limits and a transition between the two limits when
the system goes from optically thin to optically thick conditions
should introduce only limited fractional errors into
\eqntworef{eq:lout}{eq:lin}. 

The simplest estimate of the fraction of photons emitted at a given
radius that are likely to be absorbed at smaller radii comes from
finding the apparent size of the sphere where $\opticaldepth=1$ as
seen from the radius $r$, where the photons are emitted.
\begin{equation}
\Psi = 1 - \frac{1}{2} 
\left[\frac{1}{1+e^{-\opticaldepth}}\right]
\left[\frac{r_1^2}{\max(r_1^2, r^2)}\right] \, , 
\end{equation}
where $\opticaldepth$ is the optical depth from $r$ to infinity, $\opticaldepth =
\int_r^\infty \rho \kappa \, dr$, and $r_1$ is the radius where the
optical depth is unity: $1 = \int_{r1}^\infty \rho \kappa \, dr$.
Here the factor $1/(1+e^{-\opticaldepth})$ interpolates between the
correct asymptotic expressions for the case where $\opticaldepth \ll
1 \, , \, r_1 \ll r$, in which
$\Psi \simeq 1-0.5 r_1^2/r^2$ and the case where $\opticaldepth \gg 1
\, , \, r_1 \gg r$ in which $\Psi \simeq 1/2$.

\begin{figure}
  \centering
  \includegraphics[width=0.6\columnwidth]{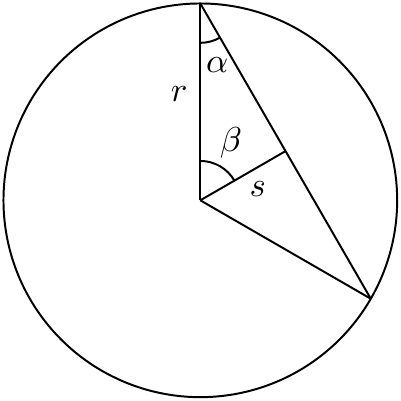}
  \caption{Geometry for calculation of optical depth in Section \ref{sec:psi}}
  \label{fig:diagram}
\end{figure}

More rigorous formulae can be derived assuming spherical symmetry and
a power-law dependence of the mean free path for absorption on the
radius.  Given the geometry shown in Figure \ref{fig:diagram}, the
optical depth as a function of the angle $\alpha$ between the
direction of the ray and the line connecting the emission point to the
center of the coordinate system is
\begin{equation}
  \opticaldepth(\alpha) = \int_0^{\pi-2\alpha} \frac{r \sin\alpha \, d\beta}
{\sin^2(\alpha + \beta) \, 
  l\left(\frac{r \sin\alpha}{\sin(\alpha+\beta)}\right)} \, ,
\label{eq:tau-raw}
\end{equation}
where $r$ is the radius of emission, $l(s)\equiv 1/\rho(r)\kappa(r)$
is the mean free path to absorption and $\alpha$, $\beta$, and $s$ are
defined by Figure \ref{fig:diagram}.  The formula gives the optical
depth for a photon to traverse the chord of the circle in Figure
\ref{fig:diagram}.  

If the mean free path as a function of radius can be approximated by a
power law
\begin{equation}
  l(s) \approx l_0 \left(\frac{s}{r}\right)^\zeta \, , \qquad \zeta > 0
\end{equation}
where $l_0$ is the local mean free path at the radius $r$ at which the
photons are emitted, then  
\eqnref{eq:tau-raw} can be written in terms of hypergeometric
functions.  However, assuming that $\zeta$ is an integer gives the simple
formulae involving elementary functions given in Table \ref{tab:taus}

\begin{table}
\centering
 \begin{minipage}{\columnwidth}
   \caption{Optical depth as a function of photon emission direction
     in the case of a power-law mean free path, where $\zeta$ is the
     power-law index.}
  \label{tab:taus}
  \begin{tabular}{@{}ll@{}}
    \hline
    $\zeta$    &  $\opticaldepth(\alpha)$ \\ 
    \hline
    0 & $(2 r/l_0) \cos \alpha   $ \\
    1 & $(2 r/l_0) \log \cot (\alpha/2) $ \\ 
    2 & $(r/l_0) (\pi-2\alpha)/ \sin \alpha$ \\
    3 & $(2 r/l_0) \cos\alpha/ \sin^2\alpha $ \\
    4 & $(r/ 2 l_0) (\pi-2\alpha+\sin 2\alpha )/ \sin^3\alpha$ \\
    \hline
  \end{tabular}
\end{minipage}
\end{table}

The transmission probability $\ptrans$ is then 
\begin{equation}
  \ptrans = \int_0^{\pi/2} \sin\alpha \, e^{-\opticaldepth(\alpha)}
  \, d\alpha \, ,
\label{eq:ptrans-integral}
\end{equation}
so that $\Psi$ can be recovered from \eqnref{eq:psi}.

The integral in \eqnref{eq:ptrans-integral} unfortunately cannot be
expressed in terms of elementary functions.  However, for a given
dependence of mean free path on radius, it is a function {\em only} of
the radius in units of the local mean free path $r/l_0$.  For
integral values of $\zeta$ the integrand consists of elementary
functions, so it can be tabulated at the beginning of a
calculation and used throughout with little computational cost.  For
values of $\zeta$ from 0 to 4 (corresponding to cases from $\rho=$
constant to $\rho \propto r^{-4}$ if the opacity is constant), the
values of $\ptrans$ and $\pabs$ do not depend sensitively on $\zeta$.

\begin{figure*}
  \includegraphics[width=0.8\textwidth]{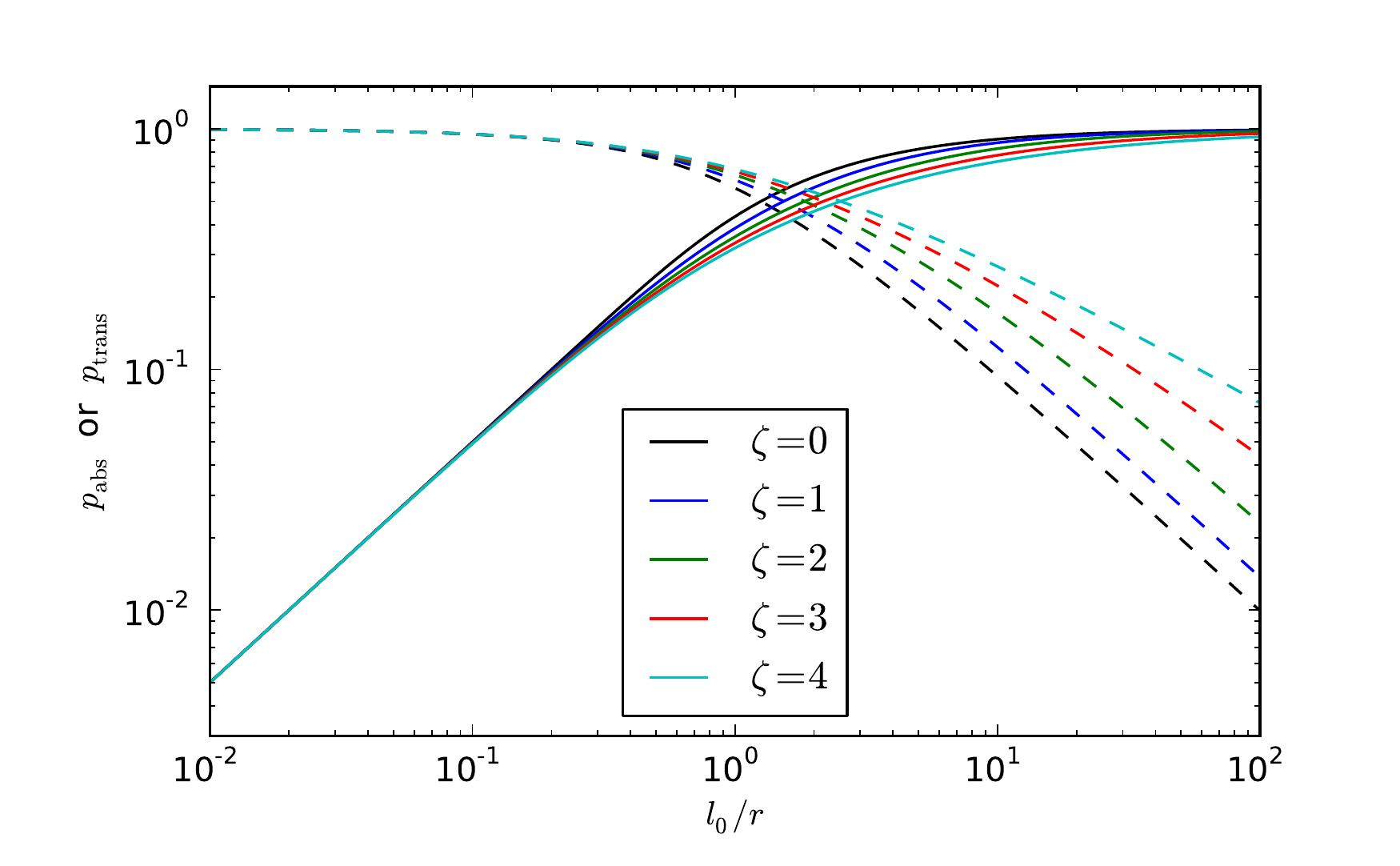}
  \caption{Probability of transmission and absorption for spherically
  symmetric systems with power-law mean free paths of index $\zeta$.
  Solid lines give
  $\pabs$, dashed lines give $\ptrans$.}
  \label{fig:psi}
\end{figure*}

Figure \ref{fig:psi} shows $\pabs$ and $\ptrans$ as the
as a function of optical depth for different values of $\zeta$.  The
precise value of $\zeta$ makes a moderate difference when the mean free
path is roughly equal to the radius of the sphere.  In the optically
thick limit, the asymptotic form of $\ptrans$ is independent of $\zeta$
and is given by
\begin{equation}
  \ptrans \approx 0.005 \left( \frac{l_0}{100 \, r} \right)\, , \quad
  \mbox{for} \quad l_0 \ll r \, .
\end{equation}
For the optically thin case, the asymptotic value of $\pabs$
depends on $\zeta$.  For values of $\zeta$ from 0 to 4, the asymptotic
form can be approximated as:
\begin{equation}
  \pabs \approx \frac{1}{100} \left[1 + 6\left(\frac{\zeta}{4}\right)^2
  \right] \left(\frac{\zeta}{4}\right)^{-1 + 0.4(\zeta/4)^3} \, , \quad
  \mbox{for} \quad l_0 \gg r \, .
  \label{eq:pabs-approx}
\end{equation}

Following the dictum discussed above that errors in $\ptrans$ are
tolerable because $\ptrans$ is always added to an order unity
constant, while the same is not true for $\pabs$ for the ingoing
radiation stream when $\pabs$ is small, then one may take the
approximation:
\begin{equation}
  \pabs = \frac{\tildepabs}{1+\tildepabs} \, ,
\end{equation}
with $\tildepabs$ given by \eqnref{eq:pabs-approx} over the whole
range of optical depths.  The value of $\Psi$ may then be recovered
using equation \ref{eq:psi}.

\label{lastpage}

\end{document}